\documentclass[smallextended,nopacs]{svjour3}

\usepackage[utf8]{inputenc}
\usepackage[T1]{fontenc}
\usepackage[english]{babel}
\usepackage{amsmath}
\usepackage{bbm}
\usepackage{nicefrac}
\usepackage{slashed}
\usepackage{pstricks}
\usepackage{graphicx}
\usepackage{hyperref}
\usepackage{leftidx}
\usepackage{environ}
\usepackage{mathtools}
\usepackage{xspace}
\usepackage{suffix}
\usepackage{url}
\usepackage{slashed}
\usepackage{scalerel,stackengine}
\usepackage{array}
\usepackage{cuted}
\usepackage{isotope}
\usepackage[margin=1in]{geometry}

\newcommand{\ie}{\textit{i.e.}\xspace}
\newcommand{\eg}{\textit{e.g.}\xspace}
\newcommand{\cf}{\textit{cf.}\xspace}
\newcommand{\apriori}{\textit{a priori}\xspace}
\WithSuffix\newcommand\apriori*{\textit{a-priori}\xspace}

\newcommand{\via}{\textit{via}\xspace}

\newcommand{\viceversa}{\textit{vice versa}\xspace}

\newcommand{\mathspace}{\ \ }
\newcommand{\mathtext}[1]{\mathspace\text{#1}\mathspace}

\newcommand{\fm}{\ensuremath{\mathrm{fm}}}

\newcommand{\vecr}{\mathbf{r}}
\newcommand{\vecx}{\mathbf{x}}

\newcommand{\vZero}{\mathbf{0}}
\newcommand{\vNabla}{\boldsymbol{\nabla}}

\newcommand{\dd}{\mathrm{d}}

\newcommand{\del}[1]{\frac{\partial}{\partial #1}}
\newcommand{\deli}[2]{\frac{\partial^{#2}}{\partial #1^{#2}}}

\newcommand{\ii}{\mathrm{i}}
\newcommand{\ee}{\mathrm{e}}

\newcommand{\OO}{\mathcal{O}}

\newcommand{\sgn}{\mathrm{sgn}}

\newcommand{\bra}[1]{\langle #1|}

\newcommand{\ket}[1]{|#1\rangle}

\newcommand{\braket}[2]{\langle #1|#2\rangle}

\newcommand{\mbraket}[3]{\langle #1|#2|#3\rangle}

\newcommand{\abs}[1]{\left|#1\right|}

\newcolumntype{C}{>{$}p{1em}<{$}}

\newcommand*\rvec[1]%
{\ensuremath{\overset{\smash{\raisebox{-1.5pt}{\tiny$\rightarrow$}}}{#1}}}
\newcommand*\lvec[1]%
{\ensuremath{\overset{\smash{\raisebox{-1.5pt}{\tiny$\leftarrow$}}}{#1}}}

\newcommand{\MeV}{\ensuremath{\mathrm{MeV}}}

\NewEnviron{subalign}[1][]{%
\begin{subequations}\begin{align}
  \BODY
\end{align}\label{#1}\end{subequations}
}

\NewEnviron{spliteq}{%
\begin{equation}\begin{split}
  \BODY
\end{split}\end{equation}
}

\stackMath
\newcommand\reallywidehat[1]{%
\savestack{\tmpbox}{\stretchto{%
  \scaleto{%
    \scalerel*[\widthof{\ensuremath{#1}}]{\kern-.6pt\bigwedge\kern-.6pt}%
    {\rule[-\textheight/2]{1ex}{\textheight}}
  }{\textheight}%
}{0.5ex}}%
\stackon[1pt]{#1}{\tmpbox}%
}

\newcommand{\ph}{{\phantom{0}}}

\newcommand{\cS}{\mathcal{S}}

\newcommand{\dvrsum}[2]{\sum\limits_{#1={-}#2/2}^{#2/2-1}}

\begin{document}

\title{Few-body bound states and resonances in finite volume}
\titlerunning{Few-body bound states and resonances in finite volume}

\author{Sebastian König}
\authorrunning{Sebastian König}

\institute{%
Institut für Kernphysik,
Technische Universität Darmstadt,
64289 Darmstadt, Germany
\and
ExtreMe Matter Institute EMMI,
GSI Helmholtzzentrum für Schwerionenforschung GmbH,
64291 Darmstadt, Germany
\and
Department of Physics,
North Carolina State University,
Raleigh, NC 27695, USA\\
\email{skoenig@ncsu.edu}
}

\date{\today}

\maketitle

\begin{abstract}
Since the pioneering work of Lüscher in the 1980s it is well known that
considering quantum systems in finite volume, specifically, finite periodic
boxes, can be used as a powerful computational tool to extract physical
observables.
While this formalism has been worked out in great detail in the
two-body sector, much effort is currently being invested into deriving
analogous relations for systems with more constituents.
This work is relevant not only for nuclear physics, where lattice methods are
now able to calculate few- and many-nucleon states, but also for other fields
such as simulations of cold atoms.
This article discusses recent progress regarding the extraction
of few-body bound-state and resonance properties from finite-volume
calculations of systems with an arbitrary number of constituents.
\end{abstract}

\section{Introduction}
\label{sec:Intro}

It is well known from the pioneering work of
Lüscher~\cite{Luscher:1985dn,Luscher:1986pf,Luscher:1990ux}
that simulating physical systems in a finite volume can be used as a tool to
extract physical properties.
The bound-state relation connects the finite-volume correction of binding
energies to the asymptotic properties of the two-particle wavefunction, whereas
for elastic scattering physical scattering parameters are encoded in the volume
dependence of discrete energy levels.
Resonances, \ie, short-lived, unstable states, are manifest in this discrete
spectrum as avoided crossing of energy levels as the size of the volume is
varied~\cite{Wiese:1988qy,Luscher:1991cf,Rummukainen:1995vs}.

All this work is based on the fact that the physical S-matrix governs the volume
dependence of energy levels and is widely used in Lattice QCD (LQCD).
It has been extended in several directions, including non-zero angular
momenta~\cite{Luu:2011ep,Konig:2011nz,Konig:2011ti}, moving
frames~\cite{Kim:2005gf,Rummukainen:1995vs,Bour:2011ef,Davoudi:2011md,%
Rokash:2013xda}, generalized boundary
conditions~\cite{Sachrajda:2004mi,Briceno:2013hya,Korber:2015rce,%
Cherman:2016vpt,Schuetrumpf:2016uuk}, particles with intrinsic
spin~\cite{Briceno:2014oea}, and perturbative Coulomb
corrections~\cite{Beane:2014qha}.

To date, most results have been obtained for two-body systems.
As numerical techniques, both LQCD and in particular lattice effective field
theory (LEFT)~\cite{Epelbaum:2013paa,Elhatisari:2015iga,Elhatisari:2016owd},
progress to calculate states with an increasing number of constituents,
understanding the volume dependence of more complex systems is of great
relevance.
This is particularly true for the study of few-body resonances in light of
recent efforts to observe~\cite{Kisamori:2016jie} and
calculate~\cite{Witala:1999pm,Lazauskas:2005ig,Hiyama:2016nwn,Klos:2016fdb,%
Shirokov:2016ywq,Gandolfi:2016bth,Fossez:2016dch,Deltuva:2018lug,%
Deltuva:2019mnv} few-neutron resonances in nuclear physics.

Early studies of the triton and Efimov trimers in finite
volume~\cite{Kreuzer:2010ti,Kreuzer:2012sr,Kreuzer:2013oya,Meissner:2014dea}
derived explicit results for these bound systems.
Generally, however, finite-volume three-body systems have a complicated
structure~\cite{Polejaeva:2012ut}, the understanding of which is an area of very
active current research~\cite{Hansen:2015zga,Briceno:2012rv,Hammer:2017uqm,%
Hammer:2017kms,Mai:2017bge,Doring:2018xxx,Pang:2019dfe,Culver:2019vvu,%
Briceno:2019muc,Romero-Lopez:2019qrt}.

This work goes in a more general direction and considers what can be said about
the volume dependence of systems with an arbitrary number $N$ of constituents.
Summarizing (and elaborating on) original work presented in
Refs.~\cite{Konig:2017krd,Klos:2018sen}, a general overview of the $N$-body
setup in Sec.~\ref{sec:NBodySetup} is followed by a discussion of bound
states and resonances in Secs.~\ref{sec:BoundStates} and~\ref{sec:Resonances},
respectively.
For bound states, the emphasis is on formal developments, while the
finite-volume study of few-body resonances is more exploratory to date and the
focus is therefore on an efficient numerical framework to look for such states.
A conclusion and outlook to future work is provided in Sec.~\ref{sec:Summary}.

\subsection{General setup}
\label{sec:NBodySetup}

Let $\ket\psi$ describe a nonrelativistic quantum state of $N$ particles in
$d$ spatial dimensions with masses $m_1,\cdots m_N$, using units where
$\hbar=c=1$.
The position-space wavefunction of this state $\ket\psi$ can be written as
$\psi({\vecr}_1,\cdots{\vecr}_N)$, where ${\vecr}_i$ labels the coordinate of
the $i$-th particle in the system.
$\ket\psi$ is assumed to be an eigenstate of a Hamiltonian
\begin{equation}
 \hat{H}_{1\cdots N} = \sum_{i=1}^N \hat{K}_{i}+\hat{V}_{1\cdots N} \,,
\label{eq:H-general}
\end{equation}
where $\hat{K}_{i} = {-\vNabla}^2_i/{(2m_i)}$
and potential term $\hat{V}_{1\cdots N}$ includes in general nonlocal
interactions of every kind from two-particle up to $N$-particle interactions:
\begin{multline}
 V_{1\cdots N}({\vecr}_1,\cdots{\vecr}_N ;{\vecr}'_1,\cdots{\vecr}'_N)
 = \sum_{i < j} W_{i,j}({\vecr}_i,{\vecr}_j;
 {\vecr}'_i,{\vecr}'_j) 1_{\slashed{i},\slashed{j}}  \\
 \null + \sum_{i< j< k}W_{i,j,k}({\vecr}_i,{\vecr}_j,{\vecr}_k;
 {\vecr}'_i,{\vecr}'_j,{\vecr}'_k) 1_{\slashed{i},\slashed{j},\slashed{k}}
 + \cdots \,,
\label{eq:vtot}
\end{multline}
where
\begin{equation}
 1_{\slashed{i}_1,\cdots\slashed{i}_k}
 = \prod_{j\ne{{i}_1,\cdots {i}_k}}\delta^{(d)}({\vecr}_j-{\vecr}'_j)
\end{equation}
conveniently accounts for spectator particles and the $W_{i,\cdots}$ are
integral kernels involving an increasing number of coordinates as one goes from
two-body towards three- and higher-body interactions.

Different kinds of relative coordinates will be used in the following.
For numerical calculations it is most convenient (for the implementations
discussed in this work) to work with simple relative coordinates defined as
\begin{equation}
 \vecx_i = \sum\limits_{j=1}^N U_{ij} \vecr_j \,,
\label{eq:x-i}
\end{equation}
where
\begin{equation}
 U_{ij} = \begin{cases}
  \delta_{ij}\,, &\text{for}\quad i,j < N\,, \\
  {-1}\,,        &\text{for}\quad i < N,\, j = N\,, \\
  1/N\,,         &\text{for}\quad i = N\,.
 \end{cases}
\label{eq:U-rel}
\end{equation}
That is, all particle coordinates are expressed relative to the last particle.
Note that for $i=N$ this definition includes the overall center-of-mass (c.m.)
coordinate.

Assuming the interactions to respect Galilean invariance, the c.m.\ momentum is
conserved and the c.m.\ kinetic energy decouples from the relative motion of the
$N$-particle system.
The kernels $W_{i,\cdots}$ can be expressed in terms of the $\vecx_i$, and
by rotational symmetry they depend only on absolute values of pairwise relative
distances.
For the special case of local interactions one has
\begin{equation}
 W_{i,j,\cdots}({\vecr}_i,{\vecr}_j,\cdots;{\vecr}'_i,{\vecr}'_j,\cdots)
 = V_{i,j,\cdots}\big(\{\abs{\vecx_i},\abs{\vecx_{i}-\vecx_{j}}_{i< j}\}\big)
 \prod_{k} \delta^{(d)}({\vecx}'_k-{\vecx}_k^\ph) \,.
\label{eq:V-gen-local}
\end{equation}
The notation for the arguments on the right-hand side is meant to indicate that
the $V_{i,j,\cdots}$ are functions of some subset of $\abs{\vecx_i}$ and
$\abs{\vecx_{i}-\vecx_{j}}$, which is sufficient to recover the relative
distances between all interacting particle pairs.
For a two-body system, the expression reduces to the familiar
$V_{1,2}(\abs{\vecx_1})\,\delta^{(d)}({\vecx}'_1-{\vecx}_1^\ph)$.

Throughout the rest of this work it is assumed that every interaction has finite
range, \ie, each $W_{i_1\cdots i_k}$ vanishes whenever the separation
between some pair of incoming or outgoing coordinates exceeds some finite
length.
The overall range $R$ of $\hat{V}_{1\cdots N}$ is defined as the maximum of all
the individual finite ranges.

\section{Bound states}
\label{sec:BoundStates}

Consider now an $N$-particle bound state with total c.m.\ momentum zero,
energy $E={-}B_N<0$, and wave function $\psi^B_N({\vecr}_1,\cdots{\vecr}_N)$.
The finite-volume behavior of this state, that is, the functional form of the
volume-dependent binding energy $B_N(L)$, is linked to the asymptotic properties
of the wavefunction when one of the coordinates becomes asymptotically large
while keeping the others fixed.
Without loss of generality, it suffices to consider the limit
$\abs{\vecr_1}\to\infty$.

Let $\cS$ denote the set of coordinate points $\{\vecr_1,\cdots\vecr_N\}$ where
$\vecr_1$ is separated by a distance greater than $R$ from all other
coordinates so that within $\cS$ there are no interactions coupling $\vecr_1$ to
$\vecr_2,\cdots\vecr_N$.
By the assumption of vanishing c.m.\ momentum, it suffices to consider the
reduced Hamiltonian
\begin{equation}
 \sum_{i=2}^{N} \hat{K}_i - \hat{K}^{\text{CM}}_{2\cdots N}
 + \hat{V}_{2\cdots N} + \hat{K}^{\text{rel}}_{1|N-1} \,,
\label{eq:H-reduced}
\end{equation}
where $\hat{K}^{\text{CM}}_{2\cdots N}
= {-}({\vNabla}_2+\cdots{\vNabla}_{N})^2/(2m_{2 \cdots N})$ and
\begin{equation}
\hat{K}^{\text{rel}}_{1|N-1}
 = {-}\frac{\left(m_{2\cdots N}{\vNabla}_1 - m_1{\vNabla}_{2\cdots N}\right)^2}
 {2\mu_{1|N-1}m^2_{1\cdots N}} \,.
\label{eq:K-rel-1Nm1}
\end{equation}
in position-space representation.
The above expressions involve the total mass
\begin{equation}
 m_{n\cdots N} = m_n+\cdots+m_{N}
\end{equation}
of the $n,\cdots N$ subsystem---used with $n=1$ and $n=2$ in
Eq.~\eqref{eq:K-rel-1Nm1}---and the reduced mass $\mu_{1|N-1}$ defined \via
\begin{equation}
 \frac{1}{\mu_{1|N-1}} = \frac{1}{m_1}+\frac{1}{m_{2\cdots N}} \,.
\end{equation}

Note that the first three terms in Eq.~\eqref{eq:H-reduced} constitute just the
Hamiltonian $\hat{H}_{2\cdots N}$ of the ${\{2,\cdots N\}}$ subsystem with the
c.m.\ kinetic energy removed, while the remaining part
$\hat{K}^{\text{rel}}_{1|N-1}$ describes the relative motion of particle $1$
with respect to the center of mass of the ${\{2,\cdots N\}}$ subsystem.
Within $S$ one can use completeness and separation of variables to expand
$\psi^B_N({\vecr}_1,\cdots {\vecr}_N)$ as a linear combination of products of
eigenstates of $\hat{H}_{2\cdots N}$ (with total linear momentum zero) and
eigenstates of $\hat{K}^{\text{rel}}_{1|N-1}$:
\begin{equation}
 \psi^B_N({\vecr}_1,\cdots{\vecr}_N)
 = \sum_\alpha \psi_\alpha({\vecr}_2,\cdots{\vecr}_{N})
   \,\chi_\alpha(\vecr_{1|N-1}) \,.
\label{eq:psi-B-expansion}
\end{equation}
Here ${\vecr}_{1|N-1} = {\vecr}_1
- (m_2{\vecr}_2+\cdots+m_{N}{\vecr}_{N})/{m_{2\cdots N}}$ and $\alpha$ labels
states in the spectrum of $\hat{H}_{2\cdots N}$ (the sum in
Eq.~\eqref{eq:psi-B-expansion} is understood to include an integral if the
spectrum is not entirely discrete).

The simplest scenario is given by assuming that the ground state of
$\hat{H}_{2\cdots N}$ is a bound state with energy ${-}B_{N-1}$, wavefunction
$\psi^B_{N-1}({\vecr}_2,\cdots{\vecr}_{N})$, and vanishing total orbital angular
momentum.
All these assumptions will be relaxed later in the discussion.
As $r_{1|N-1}=\abs{{\bf r}_{1|N-1}}$ becomes large, one finds that
\begin{equation}
 \psi^B_N({\vecr}_1,\cdots{\vecr}_N)
 \propto \psi^B_{N-1}({\vecr}_2,\cdots{\vecr}_{N}) \\
 \times (\kappa_{1|N-1}r_{1|N-1})^{1-d/2} \,
 K_{d/2-1}(\kappa_{1|N-1}r_{1|N-1}) + \cdots,
\label{eq:psi-B-asymptotic-1}
\end{equation}
where $K_{d/2-1}$ is a modified Bessel function of the second kind and
\begin{equation}
 \kappa_{1|N-1} = \sqrt{2\mu_{1|N-1}\big(B_{N}-B_{N-1}\big)}
\label{eq:1|N-1}
\end{equation}
is the momentum scale that characterizes this particular channel.
For the excited states of the $N{-}1$ system, indicated by the ellipses in
Eq.~\eqref{eq:psi-B-asymptotic-1}, there will be terms analogous to the one
shown explicitly, but they are exponentially suppressed compared to the leading
contribution due to the larger energy difference with $B_{N}$.

In the general case one considers the center of mass of $A$ particles being
separated from the remaining subsystem $N{-}A$.
Without loss of generality one can choose the $A$ coordinates to be
${\vecr}_1,\cdots{\vecr}_{A}$.
Following steps analogous to the case $A=1$, using separation of variables in
the region where the two clusters are separated by a distance larger than $R$
(such that there are no inter-cluster interactions) gives the $N$-particle
wavefunction as
\begin{equation}
 \psi^B_N ({\vecr}_1,\cdots{\vecr}_N)
 \propto \psi^B_{A}({\vecr}_1,\cdots{\vecr}_{A})
 \psi^B_{N-A}({\vecr}_{A+1},\cdots{\vecr}_{N})
 \times(\kappa_{ A|N-A} r_{A|N-A})^{1-d/2} \,
 K_{d/2-1}(\kappa_{A|N-A}r_{A|N-A}) \,,
\label{eq:asymptotic2}
\end{equation}
where
\begin{subalign}
 {\vecr}_{A|N-A}
 &= \frac{m_1{\vecr}_1+\cdots+ m_{A}{\vecr}_{A}}{m_1+\cdots+m_A}
 -\frac{m_{A+1}{\vecr}_{A+1}
 + \cdots + m_{N}{\vecr}_{N}}{m_{A+1}+\cdots+m_N} \,, \\
 \frac{1}{\mu_{A|N-A}}
 &= \frac{1}{m_1+\cdots+ m_A}+\frac{1}{m_{A+1}+\cdots+m_N} \,, \\
 \kappa_{A|N-A}
 &= \sqrt{2\mu_{A|N-A}(B_{N}-B_{A}-B_{N-A})} \,,
\end{subalign}
and ${-}B_{A}$ and ${-}B_{N-A}$ are the ground-state energies of the
$A$-particle and $(N{-}A)$-particle systems, respectively.
Note that the above derivation makes the simplifying assumption that both
$-B_{A}$ and $-B_{N-A}$ are the energies associated with, respectively, $A$ and
$N{-}A$-body bound states.
If instead either of these energies is associated with a continuum threshold,
then Eq.~(\ref{eq:asymptotic2}) remains correct only up to additional prefactors
that scale as inverse powers of $\kappa_{A|N-A}r_{A|N-A}$.

Removing finally also the restriction that the relative motion between clusters
have zero orbital angular momentum, the most general asymptotic wavefunction
for the relative motion of the two clusters has the form
\begin{equation}
 (\kappa_{ A|N-A} r_{A|N-A})^{1-d/2}\sum_{\mathbf{L}} \gamma_{\mathbf{L}}
 Y_{\mathbf{L}}(\hat{{\vecr}}_{A|N-A})
 \times K_{\ell+d/2-1}(\kappa_{A|N-A}r_{A|N-A}) \,,
\label{eq:ANC1}
\end{equation}

where $Y_{\bf{L}}$ denotes the $d$-dimensional hyperspherical harmonics for spin
representation $\ell$ (the top-level hyperspherical quantum number not otherwise
indicated explicitly, see for example Ref.~\cite{Hammer:2010fw} for details) and
the $\gamma_{\mathbf{L}}$ are expansion coefficients.
This is exactly the same behavior as found in two-particle bound states with
nonzero angular momentum, discussed for $d=2$ and
$d=3$ in Refs.~\cite{Konig:2011nz,Konig:2011ti}.
For the one-dimensional case, $\ell=\mathbf{L}=0$ and $\ell = \mathbf{L}=1$
correspond to even and odd parity, respectively, with the $d=1$ hyperspherical
harmonic being simply unity for even parity, while for odd parity it is an odd
step function.

Let now $B_N(L)$ denote the binding energy of the $N$-body state of interest in
a cubic periodic box of length $L$, and $B_N = B_N(\infty)$.
Then the finite volume correction to the binding energy is
\begin{equation}
 \Delta B_N(L) = B_N(L) - B_N \,,
\end{equation}
and following steps analogous to
Refs.~\cite{Luscher:1985dn,Konig:2011nz,Konig:2011ti,Meissner:2014dea} yields
that in general $\Delta B_N(L)$ receives contributions from every possible
breakup channel.
However, if the $N$-particle system can be subdivided as an $A$-particle bound
state and $(N{-}A)$-particle bound state in a relative $\ell = 0$ state, then
from the asymptotic behavior of the wavefunction derived above it follows that
the leading contribution to $\Delta B_N(L)$ is proportional to
\begin{equation}
 (\kappa_{ A|N-A}L)^{1-d/2} \, K_{d/2-1}(\kappa_{ A|N-A}L) \,.
\label{eq:L-dependence}
\end{equation}
In addition to this there are also terms that have a larger exponential
suppression, starting at $\OO\big(\ee^{-{\sqrt2\kappa L}}\big)$ for $d\geq2$,
and at $\OO\big(\ee^{-{2\kappa L}}\big)$ for $d=1$.
These can be safely neglected except possibly at very small $L$.
If the two bound states have orbital angular momentum $\ell > 0$, then the
finite volume correction has the same dependence as in
Eq.~\eqref{eq:L-dependence} along with subleading terms that are suppressed
by powers of $\kappa_{A|N-A}L$.
The functional form of these terms is exactly the same as that
derived for the $N=2$ case in Refs.~\cite{Konig:2011nz,Konig:2011ti}, with the
sign of $\Delta B_N(L)$ oscillating with even and odd $\ell$.

For the case that either or both the $A$-particle ground state and the
$(N{-}A)$-particle ground state are continuum states, the exponential
dependence will be the same, except that there is an additional power law
factor of $P(\kappa_{A|N-A}L)$ due to the integration over continuum states,
\begin{equation}
 (\kappa_{ A|N-A}L)^{1-d/2} K_{d/2-1}(\kappa_{ A|N-A}L)P(\kappa_{A|N-A}L) \,.
\label{eq:L-dependence_continuum}
\end{equation}
The functional form for this power law factor $P(\kappa_{A|N-A}L)$ is
currently not known, except for a few analytically solvable examples considered
in Ref.~\cite{Konig:2017krd}.

\subsection{Numerical implementation}
\label{sec:SimpleDiscretization}

The finite-volume behavior derived above can be verified by numerical
calculations.
The most straightforward way to do this is to discretize the
Hamiltonian~\eqref{eq:H-general} on a spatial lattice.
To factor out the overall center-of-mass motion from the beginning, this can
be done using directly the relative coordinates $\vecx_i$ defined in
Eq.~\eqref{eq:x-i}.
Using $n$ sites along each axis within a space of volume $L^d$ gives the
Hamiltonian as a matrix in an $n^{d\times(N-1)}$-dimensional vector space.

For two particles (with equal mass $m$ and reduced mass $\mu = m/2$) in one
spatial dimension the configuration-space wavefunction can be expressed in
terms of a single relative coordinate $x_1 \equiv x$.
The key step in discretizing the Hamiltonian is replacing the derivative in the
kinetic term by a finite-differences operator,
\begin{equation}
 \frac{\partial^2}{\partial x^2} \rightarrow D_2^{(k)} \,,
\end{equation}
where $k$ denotes the order of the stencil.
In the simplest case, $k=2$,
\begin{equation}
 D_2^{(2)} \psi(x)
 = \frac{1}{a^2}\big[\psi(x-a) - 2\psi(x) + \psi(x+a)\big] \,,
\label{eq:D-2}
\end{equation}
where $a=L/n$ denotes the lattice spacing.
Formally one can consider the wavefunction to be expanded in terms of functions
exactly localized on lattice sites,
\begin{equation}
 \psi(x) = \sum_{i} c_i \chi_i(x) \mathtext{,} \chi_i(x) = \delta(x - x_i) \,,
\end{equation}
for $x_i = i L/n$, $i={-}n/2,\cdots n/2-1$ denoting the $i$-th
lattice site, assuming $n$
to be even for convenience.
Local potentials are trivial to handle in this framework since one merely has
$V(x)\psi(x) \rightarrow V(x_i)\psi(x_i)$.
Overall, this procedure gives the Hamiltonian as a very sparse matrix, with
the kinetic part being tridiagonal for $k=2$ and the potential part diagonal.
In general, the discretization can more elegantly be expressed in a
second-quantized formalism, see for example Ref.~\cite{Lee:2008fa} for a
discussion in the context of lattice Monte Carlo methods.

The discretization is of course an approximation.
Most notably, it affects the dispersion relation that relates energies and
momenta.
While eigenstates of the free Hamiltonian remain plane waves on the lattice,
\begin{equation}
 \phi_j(x) = \braket{x}{\phi_j}
 = \frac{1}{\sqrt{L}}\exp(\ii p_j x) \,,
 \mathtext{,}
 p_j = \frac{2\pi j}{L} \,,
\label{eq:phi}
\end{equation}
applying Eq.~\eqref{eq:D-2} gives
\begin{equation}
 \hat{K}\phi_j(x) \to {-}D_2^{(2)} \phi_j(x)
 = \frac2{a^2}\big[\cos(a p_j)-1]\phi_j(x)
 = p_j^2 \phi_j(x) + \OO(a^2)
\label{eq:K-phi}
\end{equation}
for $\hat{K} = \hat{K}^{\text{rel}}_{\text{2-body}}$.
The use of stencils with $k>2$, determined for any $k$ as the solution of a
linear equation system~\cite{Cynar:1987aa}, pushes the corrections in
Eq.~\eqref{eq:K-phi} to higher orders in $a$.
This increases the computational cost, however, since increasing $k$ reduces the
sparsity of the Hamiltonian matrix by adding bands further away from the
diagonal.

The above procedure is implemented in a code that is made available as
Supplemental Material along with Ref.~\cite{Konig:2017krd}.
To complement the fully generic derivation, holding for any number of particles
in an arbitrary number of spatial dimensions, the implementation consists of a
generator program (conveniently written in Haskell due to the highly recursive
nature of the problem).
This generates a script (to be run with GNU Octave or
compatible software) for each desired setup.
For simplicity, this program only handles the case where all particles have
equal mass, but it is straightforward to adapt the Haskell code for
heterogeneous systems.

Periodic boundary conditions are implemented by using index maps for each
coordinate, which also makes it very easy to generate the kinetic term in the
Hamiltonian coordinates
The code supports these to $k$-th order accuracy with arbitrary even $k\geq2$.

Section~\ref{sec:DVR} discusses a more elaborate but closely related
implementation to discretize the Hamiltonian in a way that maintains the
exact continuum dispersion relations.
In that context it is also described how to include spin (or other discrete)
degrees of freedom, as well as how to explicitly construct states with definite
This is useful to simulate concrete systems of physical interest, while for
testing the bound-state volume dependence it suffices to run calculation using
the generator code provided with Ref.~\cite{Konig:2017krd}.

\subsection{Explicit numerical checks}

\begin{figure}[htb]
\centering
\includegraphics[width=0.75\textwidth]{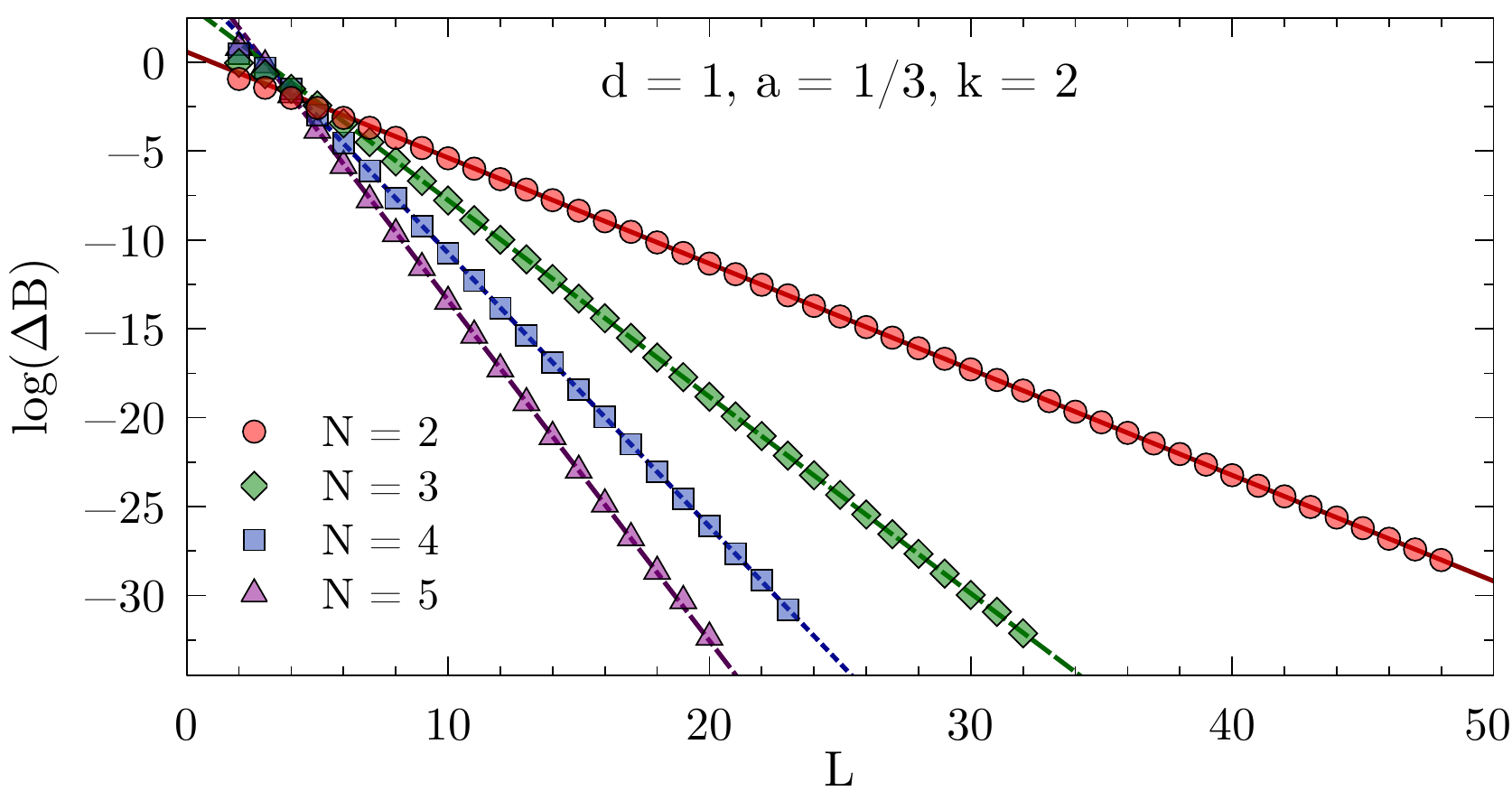}
\caption{Finite-volume energy shift for $N=2,3,4,5$ particles
interacting via a Gaussian potential ($R=1$, $V_0=-1$) in one dimension.  All
quantities are given in units of the particle mass $m=1$ (see text).}
\label{fig:En-1D-Gauss}
\end{figure}

\begin{figure}[htb]
\centering
\includegraphics[width=0.75\textwidth]{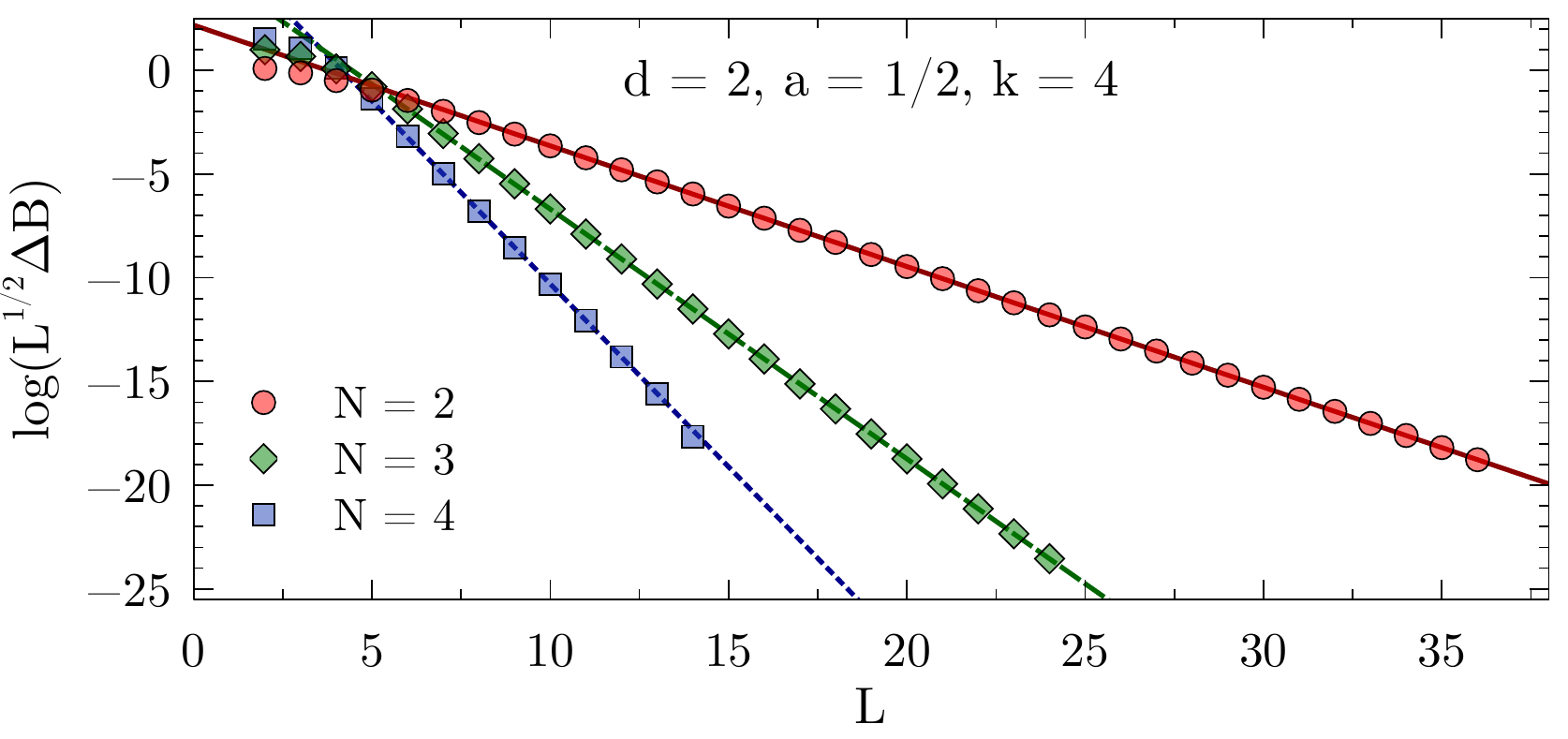}
\caption{Finite-volume energy shift for $N=2,3,4$ particles
interacting via a Gaussian potential ($R=1.5$, $V_0=-1.5$) in two dimensions.
All quantities are given in units of the particle mass $m=1$ (see text).}
\label{fig:En-2D-Gauss}
\end{figure}

\begin{figure}[htb]
\centering
\includegraphics[width=0.75\textwidth]{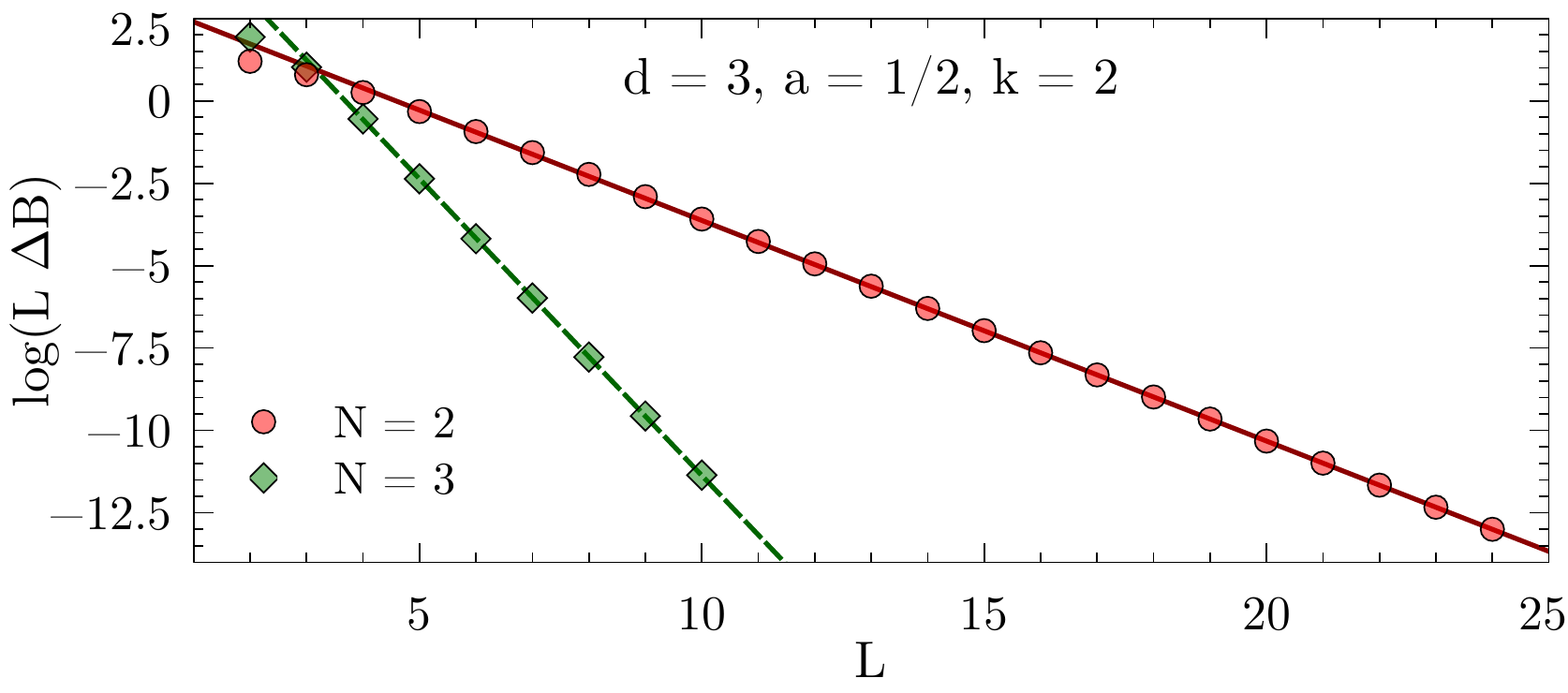}
\caption{Finite-volume energy shift for $N=2,3$ particles
interacting via a Gaussian potential ($R=1$, $V_0=-5$) in three dimensions.
All quantities are given in units of the particle mass $m=1$ (see text).}
\label{fig:En-3D-Gauss}
\end{figure}

Figures~\ref{fig:En-1D-Gauss}, \ref{fig:En-2D-Gauss}, and~\ref{fig:En-3D-Gauss}
show numerical results for, respectively, 1, 2, and 3 spatial dimensions.
These where obtained for equal-mass particles interacting \via local attractive
Gaussian potentials,
\begin{equation}
 V(r) = V_0 \exp\biggl(-\Bigl(\frac{r}{R}\Bigr)^2\biggr) \,.
\end{equation}
While these potentials do not have a strictly finite range as assumed in
the derivation of the volume dependence, their fall-off at large distances
is much faster than any expected volume dependence and therefore the relations
remain valid up negligibly small corrections.
The use of Gaussian wells instead of, \eg, strictly finite-range step
potentials has the advantage of minimizing discretization artifacts, which
are furthermore controlled in the kinetic part to calculations by using
$k=2,4,\cdots$ finite differences for the kinetic energy.
The lattice spacings $a$ for the calculations are chosen to minimize
discretization artifacts as much as possible while probing volumes large enough
to test the asymptotic behavior of the finite-volume corrections.
These calculations use natural units, which besides $\hbar=c=1$ also set the
mass to unity, $m = 1$, so that all numbers are quoted without explicit
dimension.


Expanding the Bessel function in Eq.~\eqref{eq:L-dependence} reveals that the
leading finite-volume correction has the asymptotic exponential form
\begin{equation}
 \Delta B_N(L) \propto \exp\left({-}\kappa_{A|N-A}L\right) / L^{(d-1)/2}
\label{eq:DeltaB-ND}
\end{equation}
characteristic for bound states.
This form can be easily identified by plotting the logarithm of $\Delta B_N(L)$
times $L^{(d-1)/2}$ as a function of $L$, and linear fits can be used to
extract the slopes to be compared to the expected $\kappa_{A|N{-}A}$.
The straight lines fitting the data points in the figures indicate excellent
agreement of the numerical calculation with the expected form, and
Table~\ref{tab:Results-Gauss} (which also gives the particular parameters $V_0$
and $R$ used for the Gaussian potential in each case) furthermore shows very
good quantitative agreement for the $\kappa_{A|N{-}A}$.

\begin{table}[htbp]
\centering
\begin{tabular}{cccll}
\hline\hline
$\rule{0pt}{1.2em}\phantom{x}N\phantom{x}$
& $B_N$
& $L_{\text{min}} \ldots L_{\text{max}}$
& $\hspace{1.5em}\kappa_{\text{fit}}$
& $\kappa_{1|N-1}$ \\
\hline\hline
\multicolumn{5}{c}{\rule{0pt}{1.2em}
$d=1$, $V_0 = {-}1.0$, $R = 1.0$} \\
\hline
\rule{0pt}{1.2em}%
2 & 0.356 & $20\ldots48$ & $0.59536(3)$ & 0.59625 \\
3 & 1.275 & $15\ldots32$ & $1.1062(14)$ & 1.1070 \\
4 & 2.859 & $12\ldots24$ & $1.539(3)$   & 1.541 \\
5 & 5.163 & $12\ldots20$ & $1.916(21)$  & 1.920 \\
\hline\hline
\multicolumn{5}{c}{\rule{0pt}{1.2em}
$d=2$, $V_0 = {-}1.5$, $R = 1.5$} \\
\hline
\rule{0pt}{1.2em}%
2 & 0.338 & $15\ldots36$ & $0.58195(6)$ & 0.58140 \\
3 & 1.424 & $12\ldots24$ & $1.20409(3)$ & 1.20339 \\
4 & 3.449 & $7\ldots14$  & $1.743(8)$    & 1.743 \\
\hline\hline
\multicolumn{5}{c}{\rule{0pt}{1.2em}
$d=3$, $V_0 = {-}5.0$, $R=1.0$} \\
\hline
\rule{0pt}{1.2em}%
2 & 0.449 & $15\ldots24$ & $0.6694(2)$ & 0.6700 \\
3 & 2.916 & $4\ldots14$ & $1.798(3)$ & 1.814 \\
\hline\hline
\end{tabular}
\caption{Numerical results for local Gaussian well potentials $V(r) =
V_0\exp(-r^2/R^2)$.  All quantities are given in units of the particle mass
$m=1$ (see text).}
\label{tab:Results-Gauss}
\end{table}

\begin{figure}[htb]
\centering
\includegraphics[width=0.66\textwidth]{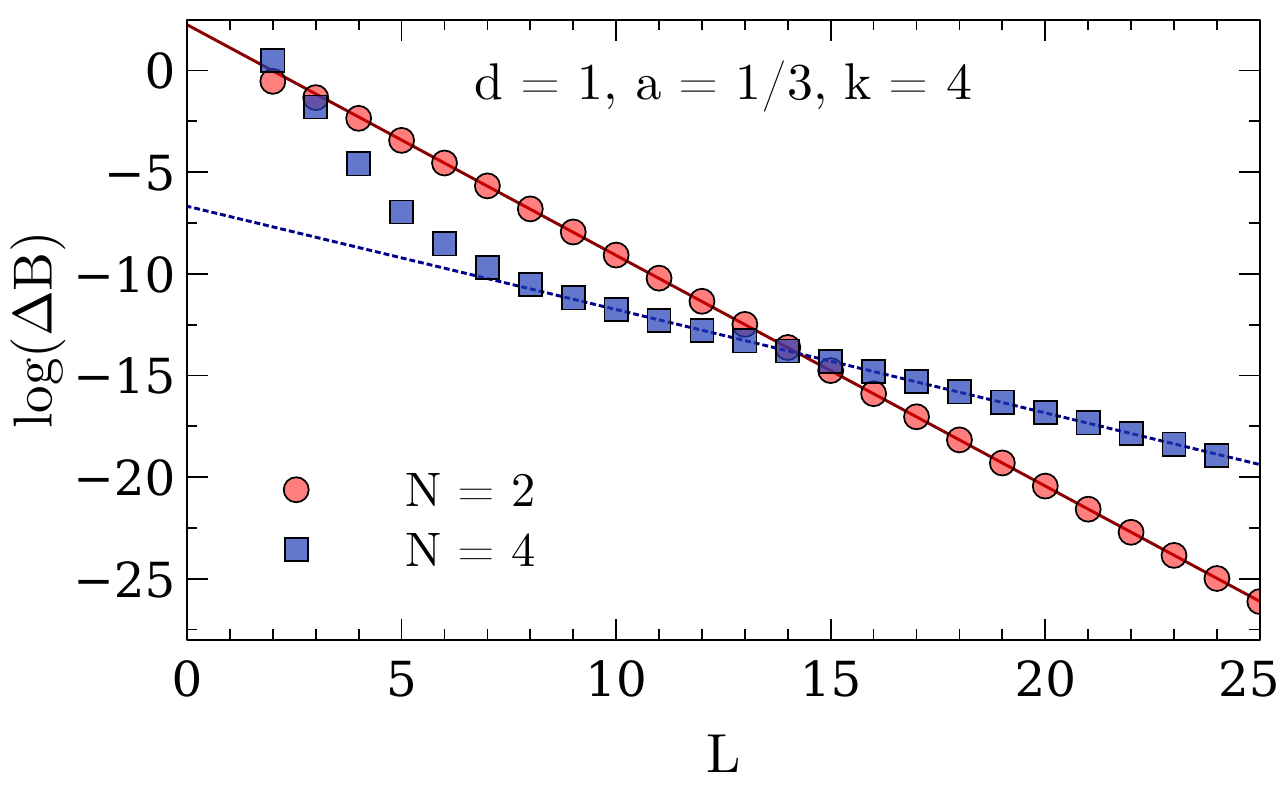}
\caption{Finite-volume energy shift for $N=2$ and $N=4$ particles
interacting via a Gaussian potentials in one dimension.
An attractive two-body potential is supplemented by a repulsive three-body one,
making the three-body system unbound, and finally by an attractive four-body
potential in order to bind that system (see text for details).
All quantities are given in units of the particle mass $m=1$.}
\label{fig:En-1D-Gauss-V234f}
\end{figure}

For all these calculations, since the interaction was chosen to be a purely
attractive two-body potential, the dominant scale is $\kappa_{1|N-1}$ because
all clusters of $N'<N$ particles are bound.
While in general one would have to heavily fine tune a two-body interaction to
create anything different from this, it is possible to introduce few-body
interactions in order to create a situation where $N=2,4$ states are bound
whereas no bound three-body state exists.
For $d=1$, a concrete example supplementing a Gaussian two-body potential with
$V_0={-}2.5$ and $R=1$, generating a two-body bound state at $E={-}1.29$,
with a repulsive local three-body force,
\begin{equation}
 V_{3}(x_1,x_2,x_{12}) = V_0^{(3)}
 \exp\Biggl({-}\biggl(\frac{x_1}{R_0^{(3)}}\biggr)^2\Biggr) \\
 \exp\Biggl(-\biggl(\frac{x_2}{R_0^{(3)}}\biggr)^2\Biggr)
 \exp\Biggl({-}\biggl(\frac{x_{12}}{R_0^{(3)}}\biggr)^2\Biggr) \,,
\label{eq:V3}
\end{equation}
where $x_{12}=|\vecx_1-\vecx_2|$, \cf Eq.~\eqref{eq:V-gen-local}.
Setting $V_0^{(3)}=10$ and $R_0^{(3)}=2$ makes the $N=3$ system unbound, which
is compensated for $N=4$ by adding an analogous short-range four-body
force---using products of Gaussians in all relative pair coordinates---with
$V_0^{(4)} = {-}24$ and $R_0^{(4)}$ to obtain a four-body bound state at
$E={-}2.71$.
The volume dependence for this system is shown in
Fig.~\ref{fig:En-1D-Gauss-V234f}.
While at small volumes the $N=4$ behavior is complicated (likely determined by a
$2+1+1$ channel), a clear linear behavior (on the appropriate log scale) is
observed asymptotically, and the extracted slope $0.508(2)$ is in excellent
agreement with $\kappa_{2|2} = 0.502$ (considering that the quoted uncertainty
is obtained from the linear fit alone).

The excellent agreement of the numerical results with the theoretical
expectation, and the fact that the volume dependence is dominated by only two
parameters, $\kappa_{A|N{-}A}$ and the proportionality factor not shown
explicitly in Eq.~\eqref{eq:DeltaB-ND}, establishes that robust extrapolations to
infinite volume can be obtained from a small set of small volumes.
Beyond that fact, knowing the functional form of the volume dependence is useful
in a more direct way because the proportionality factor is directly related to
the asymptotic normalization coefficient (ANC) associated with the $A+(N{-}A)$
threshold.
ANCs play an important role for low-energy capture processes that govern
nucleosynthesis in stellar
environments~\cite{Xu:1994zz,Capel:2013zka,Zhang:2014zsa,Hammer:2017tjm} and
are notoriously difficult to extract experimentally due to the dominance
of the Coulomb repulsion at low energies.

In the limit where separation distance $r_{A|N{-}A}$ between the two clusters is
large, the normalized $N$-body wavefunction is a product of normalized $A$-body
and $(N{-}A)$-body wavefunctions times the relative wavefunction as written
in Eq.~\eqref{eq:ANC1}.
The ANC is then the coefficient $\gamma_{\bf L}$ in Eq.~\eqref{eq:ANC1}, which
is shortened to just $\gamma$ in the following.
For the case of $d=2$, where for a precise determination of the ANC it is most
convenient to \emph{not} expand the Bessel functions as in
Eq.~\eqref{eq:DeltaB-ND} because that expansion discards some logarithmic
corrections in $d=2$ dimensions, it should be noted that this leads to a
definitions which slightly differs from the one used in
Ref.~\cite{Konig:2011ti}.
Numerically, the relative wavefunction can be obtained by calculating the ratio
\begin{equation}
 \left(\frac{\bra {\Psi^B_N} O_A(\vecr_{A|N-A}) O_{N-A}(\vZero) \ket
 {\Psi^B_N}}
 {\bra {\Psi^B_A} O_A(\vZero) \ket {\Psi^B_{A}} \bra {\Psi^B_{N-A}}
 O_{N-A}(\vZero) \ket {\Psi^B_{N-A}}}\right)^{\!1/2}
\label{eq:Psi-O}
\end{equation}
for some localized $A$-body and $(N{-}A)$-body operators $O_{A}(\vecr)$,
$O_{N-A}(\vZero)$.
The result of this determination can then be compared to the
the asymptotic form as given in Eq.~\eqref{eq:ANC1}, taking into account
additional copies due to the periodic boundary conditions.
The ratio gives the magnitude of the ANC, denoted by
$\abs{\gamma}_{\text{WF}}$ to indicate the determination directly from the
wavefunction.

In addition, the ANC can be obtained in a completely different way using the
finite-volume correction $\Delta B_N(L)$.
By combining the $N$-body results summarized here with the derivations in
Refs.~\cite{Luscher:1985dn,Konig:2011nz,Konig:2011ti}, one finds that $\Delta
B_N(L)$ equals
\begin{equation}
 \frac{(-1)^{\ell+1} \sqrt{\tfrac{2}{\pi}}
 f(d)\abs{\gamma}^2}{\mu_{A|N-A}}
 \kappa^{2-d/2}_{A|N-A}L^{1-d/2} K_{d/2-1}(\kappa_{A|N-A}L) \,,
\label{eq:ANC-FV}
\end{equation}
plus corrections that are exponentially suppressed.
This relation follows directly from defining the ANC in terms of the asymptotic
radial wavefunction, which for a cluster separation $r_{A|N-A}$ large compared
to the range of the interaction is universally given by
\begin{equation}
 \psi_{\text{asympt}}(r_{A|N-A})
 = \gamma \, \sqrt{\frac{2\kappa_{A|N-A}}{\pi}}
 (r_{A|N-A})^{1-d/2}
 K_{d/2-1}(\kappa_{A|N-A}r_{A|N-A}) \, Y(d) \,,
\label{eq:ANC-psi}
\end{equation}
where $Y(d)$ accounts for the angular normalization in $d$ spatial dimensions.
For $d=3$, where $Y(3) = 1/\sqrt{4\pi}$, the convention in
Eq.~\eqref{eq:ANC-psi} reproduces the canonical form
\begin{equation}
 \gamma\exp({-}\kappa_{A|N-A}r_{A|N-A})/r_{A|N-A}
\end{equation}
for a two-cluster S-wave state.
For $d=1$ one has $Y(1) = 1$ and the asymptotic form is simply
$\gamma$ $\times\exp({-}\kappa_{A|N-A}r_{A|N-A})$, while as already stated
for $d=2$ it is more natural to define the ANC directly in terms of the
modified Bessel function that does not fully reduce to a simple exponential
in this case.
The function $f(d)$ captures these conventional differences and takes values
$f(1)=2$, $f(2)=\sqrt{8/\pi}$, and $f(3)=3$.
For $d=3$, $\Delta B_N(L)$ is averaged over all $2\ell +1$ elements of a given
angular momentum $\ell$ multiplet, while for $d=2$ the average is taken over
symmetric and antisymmetric combinations of $\mathbf{L} = \pm\ell$ for even
$\ell$~\cite{Konig:2011ti}.
The result of this ANC extraction, using fits of Eq.~\eqref{eq:ANC-FV} to the
data shown in Figs.~\ref{fig:En-1D-Gauss}, \ref{fig:En-2D-Gauss},
\ref{fig:En-3D-Gauss}, is denoted as $\abs{\gamma}_{\text{FV}}$.
Note that if there are several different ways to partition the $N$-particle
system into clusters with same $\kappa_{A|N{-}A}$ value, then there will
contributions to the finite-volume correction from each channel.

Using the same Gaussian well potentials as discussed previously,
results for $\abs{\gamma}_{\text{FV}}$ and $\abs{\gamma}_{\text{WF}}$ are
shown in Table~\ref{tab:Results-Gauss-ANC}.
This analysis used Eq.~\eqref{eq:Psi-O} with the operator $O_1$ equal to the
single particle density and $O_{N-1}$ equal to the $(N-1)$-body density on a
single lattice site, with all quantities extracted at the same finite volume.
As seen in Table~\ref{tab:Results-Gauss-ANC}, the two methods for extracting
the ANCs are in excellent agreement.
The technique therefore provides a strikingly simple and robust way to extract
ANCs, which will be of great practical relevance once an extension of the
finite-volume formalism to include the Coulomb force is available.
Finally, it is worth noting that with $\abs{\gamma}_{\text{WF}}$ extracted from
a single volume (assuming one is using a method that gives access to the
wavefunction), one can in fact determine $B_N(L)$ from a single-volume
calculation.
This can be relevant in practice for cases where calculations at
multiple volumes are prohibitively expensive.

\begin{table}[htbp]
\centering
\begin{tabular}{ccccc}
\hline\hline
$\rule{0pt}{1.2em}\phantom{x}N\phantom{x}$
& $B_N$
& $L_{\text{max}}$
& $\abs{\gamma}_{\text{FV}}$
& $\abs{\gamma}_{\text{WF}}$ \\
\hline\hline
\multicolumn{5}{c}{\rule{0pt}{1.2em}
$d=1$, $V_0 = {-}1.0$, $R = 1.0$} \\
\hline
\rule{0pt}{1.2em}%
2 & 0.356 & $48$ & $0.8652(4)$ & $0.8627(4)$ \\
3 & 1.275 & $32$ & $1.650(27)$ & $1.638(16)$ \\
4 & 2.859 & $24$ & $2.54(6)$   & $2.56(8) $ \\
5 & 5.163 & $20$ & $3.65(62)$  & $3.63(18)$ \\
\hline\hline
\multicolumn{5}{c}{\rule{0pt}{1.2em}
$d=2$, $V_0 = {-}1.5$, $R = 1.5$} \\
\hline
\rule{0pt}{1.2em}%
2 & 0.338 & $36$ & $1.923(2)$ & $1.921(9)$ \\
3 & 1.424 & $24$ & $5.204(4)$ & $5.24(2)$ \\
4 & 3.449 & $14$ & $11.2(4)$  & $10.99(4)$ \\
\hline\hline
\multicolumn{5}{c}{\rule{0pt}{1.2em}
$d=3$, $V_0 = {-}5.0$, $R=1.0$} \\
\hline
\rule{0pt}{1.2em}%
2 & 0.449 & $24$ & $1.891(3)$ & $1.89(1)$ \\
3 & 2.916 & $14$ & $7.459(97)$ & $7.83(11)$ \\
\hline\hline
\end{tabular}
\caption{Extracted ANCs for local Gaussian well potentials $V(r) =
V_0\exp(-r^2/R^2)$.
All quantities are given in units of the particle mass $m=1$.}
\label{tab:Results-Gauss-ANC}
\end{table}

\section{Resonances}
\label{sec:Resonances}

All levels in the discrete finite-volume energy spectrum that are not bound
states characterized by the asymptotic exponential behavior discussed in
Sec.~\ref{sec:BoundStates} have a power-law volume dependence.
The Lüscher formalism used to extract infinite-volume scattering observables is
based on analyzing how these levels are shifted by the interaction among the
particles compared to the free (non-interacting) energy
levels~\cite{Luscher:1986pf,Luscher:1990ux}.
As mentioned in the introduction, resonance states do not correspond to
individual energy levels at finite volume, but instead are manifest as
(sequences of) avoided crossings within the power-law spectrum.
This is well established for two-body systems~\cite{Wiese:1988qy,%
Luscher:1991cf,Rummukainen:1995vs}, whereas Ref.~\cite{Klos:2018sen} showed that
this result carries over to the few-body sector, thereby establishing
finite-volume calculations as a theoretical tool to discover resonances which
can be interpreted as metastable states of $N\geq3$ constituents.
These results, and in particular the ``discrete variable representation (DVR)''
used as numerical method for these calculations, are discussed in the following.

\subsection{Discrete variable representation}
\label{sec:DVR}

The starting point for the DVR construction used here are plane-wave states
$\phi_j(x)$, where $j={-}n/2,\cdots$ $\cdots n/2-1$ for $n>2$ even, defined in
Eq.~\eqref{eq:phi}, where $x$ at this point denotes a single relative coordinate
in one dimension.
It is clear that any periodic solution of the Schrödinger equation can be
expanded in terms of these states, and this expansion becomes exact
for $n\to\infty$.

\begin{figure}[tbhp]
\centering
\begin{minipage}{0.5\textwidth}
\centering
\includegraphics[width=0.9\textwidth]{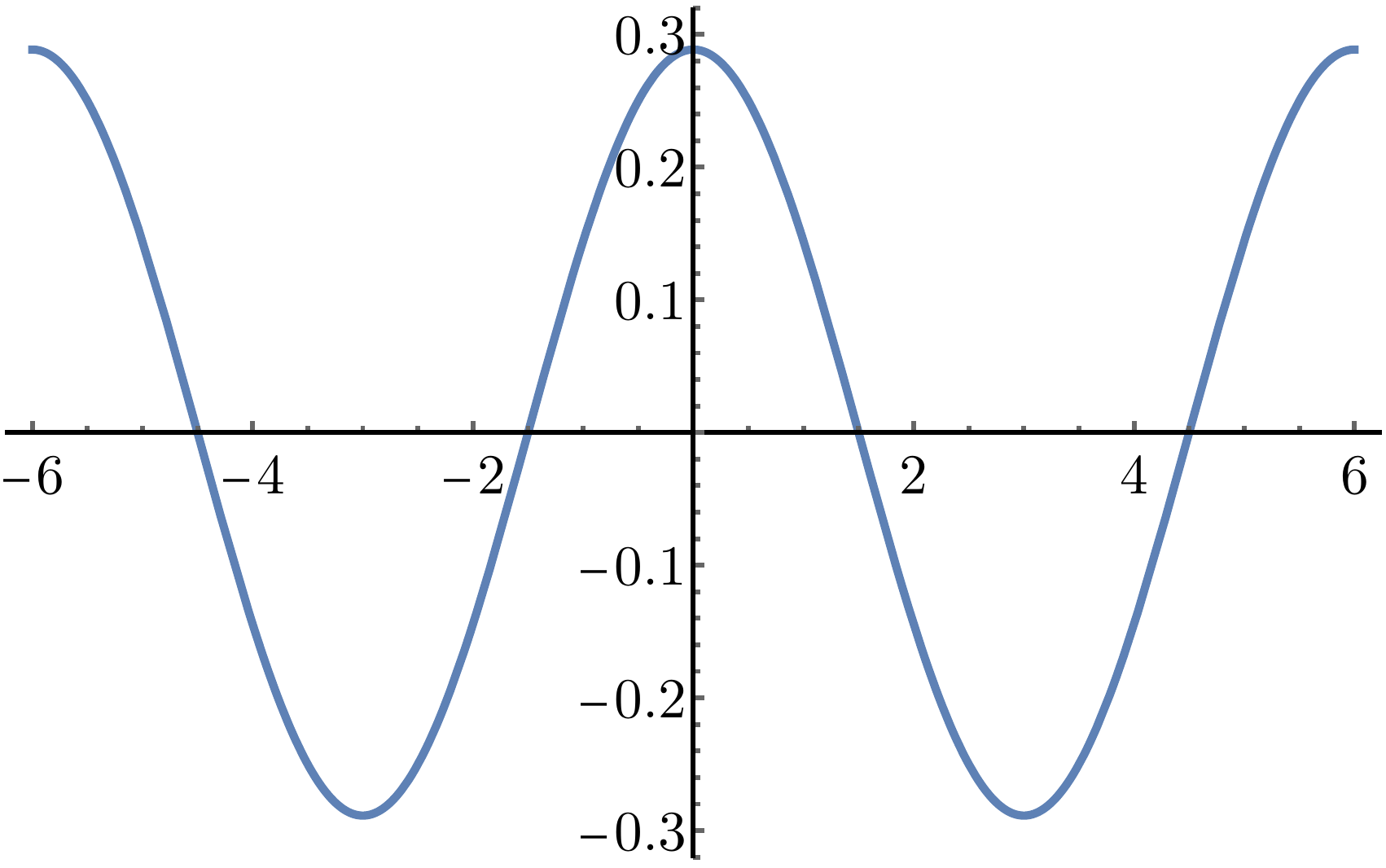}
\end{minipage}%
\begin{minipage}{0.5\textwidth}
\centering
\includegraphics[width=0.9\textwidth]{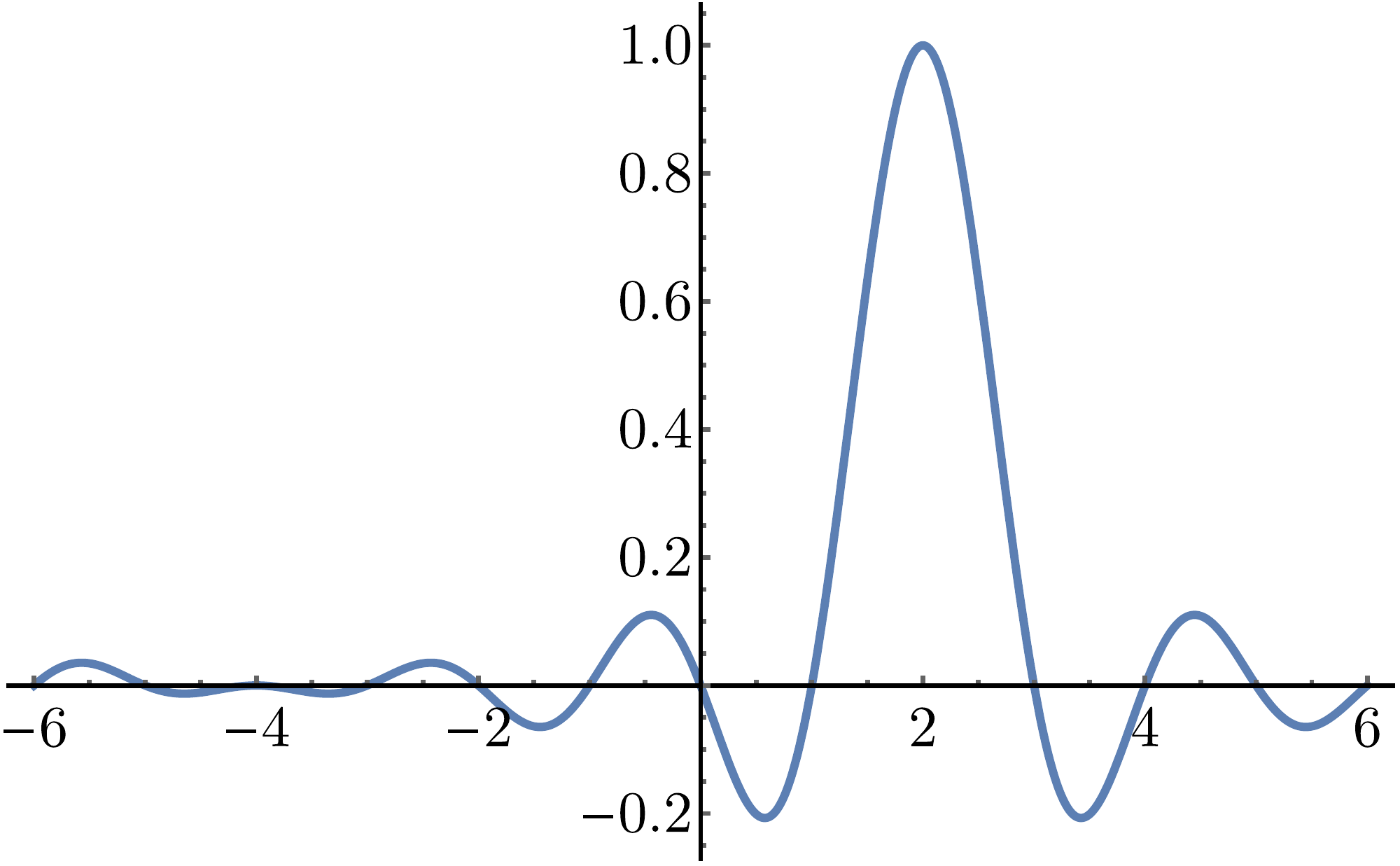}
\end{minipage}
\caption{%
Plane-wave (left) and DVR states (right) for a single relative coordinate $x$
in one dimension.  What is plotted are the real parts of the wavefunctions.}
\label{fig:dvr_states}
\end{figure}

Following the general construction described in Ref.~\cite{Groenenboom:2001web},
consider now pairs $(x_k, w_k)$ of quadrature points $x_k$ and associated
weights $w_k$ such that
\begin{equation}
 \dvrsum{k}{n} w_k\,\phi_i^*(x_k) \phi_j(x_k) = \delta_{ij} \,.
\label{eq:phi-orth}
\end{equation}
For the plane-wave states~\eqref{eq:phi}, this is satisfied by an equidistant
mesh with constant weight:
\begin{equation}
 x_k = \frac{L}{n}k
 \mathtext{,}
 w_k = \frac{L}{n} \,\forall k \,.
\label{eq:xk-wk}
\end{equation}
Equipped with this one can define matrices
\begin{equation}
 \mathcal{U}_{ki} = \sqrt{w_k} \phi_i(x_k) \,,
\label{eq:U-DVR}
\end{equation}
and these matrices are unitary by Eq.~\eqref{eq:phi-orth}.
The DVR basis functions $\psi_k(x)$ are defined by rotating the original
plane-wave basis with $\mathcal{U}^*$, where the asterisk denotes complex
conjugation:
\begin{equation}
 \psi_k(x) = \dvrsum{i}{n} \mathcal{U}^*_{ki} \phi_i(x)
\label{eq:psi-dvr}
\end{equation}
for $k = {-}n/2,\ldots,n/2-1$.
The range of indices is the same as for the original plane-wave states,
but whereas in Eq.~\eqref{eq:phi} they specify a momentum mode, $\psi_k(x)$ is
approximately localized at position $x_k \in [{-}L/2,L/2)$.
An example is shown in Fig.~\ref{fig:dvr_states}.

An explicit evaluation of Eq.~\eqref{eq:psi-dvr} reveals that
\begin{equation}
 \psi_k(x) = \dvrsum{i}{n} \phi_i(x-x_k)
 = \frac1{\sqrt{L}} \dvrsum{i}{n} \ee^{\ii p_i(x-x_k)} \,,
\label{eq:psi-dft}
\end{equation}
so the DVR construction can be related to a discrete Fourier transform
(DFT).\footnote{%
The DVR construction along Eqs.~\eqref{eq:phi-orth} and~\eqref{eq:psi-dvr} is
however much more general and needs not start from a plane-wave basis.
For example, Refs.~\cite{Binder:2015trg,Bansal:2017pwn} use a DVR basis built
out of harmonic oscillator states in momentum space.}
Indeed, as seen in the right panel of Fig.~\ref{fig:dvr_states}, $\psi_k$
approximates a delta function centered at $x_k$, and from the definitions one
finds that
\begin{equation}
 \psi_k(x_j) = \frac{1}{\sqrt{w_k}} \delta_{kj} \,.
 \label{eq:DVR-delta}
\end{equation}
This establishes a close relation to the discretization discussed in
Sec.~\ref{sec:SimpleDiscretization}.
Within the space of DVR states, the dispersion relation is exact (formally one
can think of achieving this with an infinite-order finite-difference method for
the derivative), and in line with this the kinetic energy is given by a dense
matrix:
\begin{equation}
 \mbraket{\psi_k}{\hat{K}}{\psi_l}
 = \begin{cases}
  \dfrac{\pi^2 N^2}{6\mu L^2} \left(1+\dfrac{2}{n^2}\right)\,,
  &\text{for}\quad k = l\,, \\[10pt]
  \dfrac{({-}1)^{k-l} \pi^2}{\mu L^2 \sin^2\big(\pi(k-l)/n\big)}\,,
  &\text{otherwise}\,.
 \end{cases}
\label{eq:T-DVR-1D}
\end{equation}
This is more computationally demanding that the band-diagonal structure
obtained from a simple finite-difference discretization, but still
the matrix elements are known explicitly for any index pair $(k,l)$.
Moreover, for $d>1$ the matrix becomes sparse, as will be seen below.
Alternatively, as pointed out in Ref.~\cite{Bulgac:2013mz}, one can exploit the
relation of the plane-wave based DVR to the DFT and evaluate the kinetic energy
in momentum space.
This involves transforming to the original plane-wave basis~\eqref{eq:phi},
applying
\begin{equation}
 \hat{K}\ket{\phi_i} = \frac{p_i^2}{2\mu}\ket{\phi_i} \,,
\end{equation}
and then transforming back.

Importantly, evaluation of (local) potential matrix elements is just as simple
as with the discretization of Sec.~\ref{sec:SimpleDiscretization}.
From Eq.~\eqref{eq:DVR-delta} it follows that
\begin{spliteq}
 \mbraket{\psi_k}{\hat{V}}{\psi_l}
 &= \int \dd x\,\psi_k^*(x) V(x) \psi_l(x)\,, \\
 &\approx \dvrsum{i}{n} w_m\,\psi_k^*(x_i) V(x_i) \psi_l(x_i)\
 = V(x_k) \delta_{kl} \,,
\label{eq:V-DVR}
\end{spliteq}
so that the potential operator is diagonal in the DVR representation.
The approximation indicated in the second line in Eq.~\eqref{eq:V-DVR} lies in
replacing the integral by a sum, which is possible because the $(x_k,w_k)$
defined in Eq.~\eqref{eq:xk-wk} constitute the mesh points and weights of a
trapezoidal quadrature rule.\footnote{Note that for this identification it is
important that the points ${-}L/2$ and $L/2$ are identified through the periodic
boundary condition because otherwise the weight $w_{{-}n/2}$ would be
incorrect.}
While not very accurate in general, this quadrature rule is highly efficient
for integrating periodic functions.

\subsubsection{General construction}

The construction is straightforward to generalize to the case of an arbitrary
number of particles $N$ and spatial dimensions $d$, starting from product states
of $(N-1)\times d$ plane waves, one for each relative-coordinate component.
The transformation matrices and DVR basis functions are defined \via tensor
products, and DVR states are labeled by a collection of $(N-1)\times d$ indices.
Using the short-hand notation $\ket{\psi_k} = \ket{k}$, a general state is
written as
\begin{equation}
 \ket{s}
 = \ket{
  (k_{{1,1}},\cdots k_{{1,d}}),\cdots,(k_{{N-1,1}},\cdots k_{{N-1,d}});
  (\sigma_1,\cdots,\sigma_N)
 } \,,
\label{eq:s}
\end{equation}
including additional indices $\sigma_i$ to account for spin degrees of
freedom.
If the particles have spin $S$, then each $\sigma_i$, labeling the projections,
takes values from ${-}S$ to $S$.
Additional internal degrees of freedom, such as isospin, can be included in the
same way.
The space spanned by all these states $\ket{s}$ is denoted by $B$.

As already mentioned, the kinetic energy becomes a sparse matrix for $d>1$.
A one-dimensional matrix element~\eqref{eq:T-DVR-1D} enters for each component
$k_{i,c}$, multiplied by Kronecker deltas for all $c' \neq c$ and summed for all
relative coordinates $i=1,\ldots,N-1$.
Working with simple relative coordinates as defined in Eq.~\eqref{eq:x-i}
implies that the general kinetic energy operator,
\begin{equation}
 \hat{K}^{\text{rel}}_{\text{$N$-body}}
 = {-}\frac{1}{2\mu} \sum\limits_{i=1}^{N-1}\sum\limits_{j=1}^i
 \del{x_i}\del{x_j} \,,
\label{eq:T-rel}
\end{equation}
contains mixed (non-diagonal) terms, \eg,
\begin{equation}
 \hat{K}^{\text{rel}}_{\text{3-body}}
 = -\frac{1}{2\mu}\biggl(\deli{x_1}{2}+\deli{x_2}{2}+\del{x_1}\del{x_2}\biggr)
\end{equation}
for three particles in one dimension the kinetic-energy matrix elements are
given by
\begin{equation}
 \mbraket{k_{1}k_{2}}{\hat{K}^{\text{rel}}_{\text{3-body}}}{l_{1}l_{2}}
 = \mbraket{k_1}{\hat{K}^{\text{rel}}_{x_1}}{l_1}\delta_{k_2 l_2}
 + \mbraket{k_2}{\hat{K}^{\text{rel}}_{x_2}}{l_2}\delta_{k_1 l_1}
 + \mbraket{k_{1}k_{2}}{\hat{K}^{\text{rel}}_{x_1-x_2}}{l_{1}l_{2}} \,,
\end{equation}
where the first two matrix elements on the right-hand side are given in
Eq.~\eqref{eq:T-DVR-1D} and the last term is a special case of the general
mixed-derivative matrix element
\begin{equation}
 \mbraket{k_{i}k_{j}}{\hat{K}^{\text{rel}}_{x_i-x_j}}{l_{i}l_{j}}
 = {-}\frac{1}{2\mu}
 \big[\mbraket{k_i}{\partial_i}{l_i} \mbraket{k_j}{\partial_j}{l_j}\big]
\end{equation}
with~\cite{Bilaj:2017bsc}
\begin{equation}
 \mbraket{k}{\partial}{l} = \begin{cases}
  {-}\ii\dfrac{\pi}{L}\,,
  &\text{for}\quad k = l\,, \\
  \dfrac{\pi}{L}
  \dfrac{({-}1)^{k-l}\exp\!\left({-}\ii\dfrac{\pi(k-l)}{n}\right)}
  {\sin\!\left(\dfrac{\pi(k-l)}{n}\right)}\,,
  &\text{otherwise}\,.
 \end{cases}
\end{equation}
As for the diagonal terms, for a general state $\ket{s}$ such terms are summed
over for all pairs of relative coordinates and spatial components $c$, including
Kronecker deltas for $c' \neq c$.

\subsubsection{Reduction by symmetry}
\label{sec:DVR-Symmetry}

In the form introduced above, the DVR basis includes states with many different
symmetry properties.
To focus on a particular sector of interest, there are different possibilities
to single out subspaces of the full Hilbert space which the truncated basis
approximates.
The most direct approach explicitly constructs linear combinations of states
with the desired properties.
It is a feature of the DVR basis that for some important cases this construction
can be carried out with great efficiency.

To study systems of identical bosons (or fermions) it is necessary to consider
subspaces of states which are fully (anti-)symmetric under permutations of the
individual particles.
A convenient way to construct such states follows the method described in
Ref.~\cite{Varga:1997xga}.
While that paper considers the stochastic variational model in Jacobi
coordinates, it is straightforward to adapt the procedure for DVR states
expressed in simple relative coordinates.
The relevant steps are as follows:
\begin{enumerate}
 \item The transformation from single-particle to relative coordinates (and
\viceversa) is constructed as given in Eq.~\eqref{eq:U-rel}
\item For the $N$-particle system there are $N!$ permutations, constituting the
symmetric group $S_N$.
A permutation $p \in S_N$ can be represented as a matrix $C(p)$ with
\begin{equation}
 C(p)_{ij} = \begin{cases}
  1\,, &\text{for}\quad j = p(i)\,, \\
  0\,, &\text{otherwise}\,,
\end{cases}
\end{equation}
acting on the single-particle coordinates $\vecr_i$.
\item The operation of $p \in S_N$ on the relative coordinates is then given by
the matrix
\begin{equation}
 C_{\text{rel}}(p) = U\,C(p)\,U^{{-1}} \,,
\end{equation}
with the last row and column of the left-hand side, corresponding to the overall
c.m.\ coordinate, discarded, so that
$C_{\text{rel}}(p)$ is an $(N{-}1)\times(N{-}1)$ matrix.
\end{enumerate}

Since the indices $k_{i,c}$ correspond directly to positions on the spatial
grid used to define the initial plane-wave states (recall that for a two-body
system in one dimension $\psi_k(x)$ is peaked at $x=x_k$), acting with
$C_{\text{rel}}(p)$ on a state $\ket{s}$ is now straightforward:
the $k_{i,c}$ are transformed according to the entries $C_{\text{rel}}(p)_{ij}$,
where for each $i$ one considers all $c=1,\ldots,d$ at once.
In other words, $C_{\text{rel}}(p)$ is expanded (by replication for each
$c$) to a matrix acting in the space of individual coordinate components.
As a final step, to maintain periodic boundary conditions, any transformed
indices that may fall outside the original range ${-}n/2,\ldots,n/2-1$ are
wrapped back into this interval by adding appropriate multiples of $n$.
Applying the permutation to the spin indices $(\sigma_1,\ldots,\sigma_N)$ is
trivial because they are given directly as an $N$-tuple.
The final result of this process for a given state $\ket{s} \in B$ and
permutation $p$ is a transformed state,
\begin{equation}
 \ket{s'} = \mathcal{C}(p) \ket{s} \in B \,,
\label{eq:C-p}
\end{equation}
where
\begin{equation}
 \mathcal{C}(p) = C_{\text{rel}}(p) \, C_{\text{spin}}(p)
\end{equation}
denotes the total permutation operator in the space of DVR states.
The statement of Eq.~\eqref{eq:C-p} is that each $p\in S_n$ acts on $B$ as a
whole by permuting the order of elements.

With this, one finds the symmetrization and antisymmetrization operators as
\begin{equation}
 \mathcal{S} = \frac{1}{n!}\sum_{p\in S_n} \mathcal{C}(p)
 \mathtext{and}
 \mathcal{A} = \frac{1}{n!}\sum_{p\in S_n} \sgn(p)\,\mathcal{C}(p) \,,
\end{equation}
where $\sgn(p) = \pm1$ denotes the parity of the permutation $p$.
Since both of these operators are projections ($\mathcal{S}^2 = \mathcal{S}$,
$\mathcal{A}^2 = \mathcal{A}$), they map the original basis $B$ onto bases
$B_\mathcal{S/A}$ of, respectively, symmetrized or antisymmetrized states, each
of which consists of linear combinations of states in $B$.

An important feature of these mappings is that each $\ket{s}\in B$ appears in at
most one state in $B_\mathcal{S}$ (for symmetrization) or $B_\mathcal{A}$ (for
antisymmetrization).
Thus, to determine $B_\mathcal{S/A}$ it suffices to apply $\mathcal{S/A}$ to
all $\ket{s}\in B$, dropping duplicates which occur when the operator is
applied to a state in $B$ that has already been generated by a previous
permutation.
This algorithm is straightforward to apply in numerical calculations and it can
be made highly efficient.
Moreover, computer memory can be saved by storing for each combined state in
$B_\mathcal{S/A}$ only the index of one state in $B$ that generates it.
This can be useful to trade memory efficiency against an increase in
computational time since the coefficients defining the combined states have to
be recalculated as needed.

\paragraph{Cubic symmetry}

While permutation symmetry and parity remain unaffected by the finite volume,
rotational symmetry (for $d>2$) is explicitly broken by the periodic box.
In $d = 3$ dimensions (to which the remaining discussion in this section will
be limited), angular momentum $\ell$ associated with spherical $SO(3)$ symmetry
is no longer a good quantum number.
Instead, one has to consider the breaking of $SO(3)$ down to a cubic subgroup
$\OO \subset SO(3)$.

This group has 24 elements and five irreducible representations $\Gamma$,
conventionally labeled $A_1$, $A_2$, $E$, $T_1$, and $T_2$.
Their dimensionalities are $1$, $1$, $2$, $3$, and $3$, respectively, and
irreducible representations $D^l$ of $SO(3)$, determining angular-momentum
multiplets in the infinite volume, are reducible with respect to $\OO$.
As a consequence, any given angular-momentum state in infinite volume can
contribute to several representations $\Gamma$.
In the cubic finite volume one finds the spectrum decomposed into multiplets
with definite $\Gamma$, where an index $\alpha=1,\ldots,\dim\Gamma$ further
labels the states within a given multiplet.

For the calculations considered here, it is desirable to select spectra by their
cubic transformation properties.
To that end, one constructs projection operators~\cite{Johnson:1982yq},
\begin{equation}
 \mathcal{P}_\Gamma
 = \frac{\dim\Gamma}{24} \sum_{R\in\OO}\,\chi_\Gamma(R) D_{n}(R) \,,
\label{eq:P_Gamma}
\end{equation}
where $\chi_\Gamma(R)$ denotes the character (tabulated in
Ref.~\cite{Johnson:1982yq}) of the cubic rotation $R$ for the irreducible
representation $\Gamma$ and $D_{n}(R)$ is the realization of the cubic rotation
in our DVR space of periodic $n$-body states.
For example, for the one-dimensional representation $\Gamma=A_1$,
$\chi_{A_1}(R)=1$ for all cubic rotations $R$, so in this case
Eq.~\eqref{eq:P_Gamma} reduces to an average over all rotated states,
analogous to how one can project onto $S$ waves ($\ell=0$) in infinite volume.
The construction of the $D_{n}(R)$ is discussed in detail in
Ref.~\cite{Klos:2018sen}.

\subsection{Numerical implementation}

The DVR scheme described above essentially amounts to constructing the
Hamiltonian $\hat{H}$ for the physical system of interest in a particular
truncated basis ($B$), which then gives rise to a finite matrix representation
$H=\hat{H}\big|_B$.
This dimension of this matrix grows with (i) the number $N$ of particles (and
their spin as well as potential other discrete degrees of freedom), (ii)
the number $d$ of spatial dimensions, and (iii) the number $n$ of DVR states
used in the calculation.
The precise scaling for particles with spin $S$ is
\begin{equation}
 \dim B = (2S+1)^N \times n^{(N-1)\times d} \,.
\end{equation}

While $N$, $S$, and $d$ are fixed by the particular physics problem, the
appropriate choice of $n$ depends on all other parameters as well as on details
of the interaction and the size $L$ of the volume.
While in principle $n$ has to be sufficiently large to converge both the kinetic
and the potential parts entering the Hamiltonian, the most important effect comes
from Eq.~\eqref{eq:V-DVR}: the representation of the (local) interaction as a
diagonal matrix rests on the quality of the quadrature that approximates the
integral.
The more peaked (or generally more ``structured'') the shape of the potential,
the larger $n$ needs to be to adequately approximate the integral, and likewise,
the larger $L$, the larger $n$ is needed since the spacing between quadrature
points increases with $L$ at fixed $n$.
In practice, a sequence of increasing $n$ has to be considered at each volume
until sufficient convergence is reached, \ie, until the change of energies with
$n$ is negligible compared to the desired precision.
It is worth noting that all symmetries discussed in Sec.~\ref{sec:DVR-Symmetry}
are by construction exact at each $n$, so there are no convergence issues as far
as this part of the calculation is concerned.
Finally, the whole setup discussed in this section could be adapted to use the
simple finite-difference discretization described in
Sec.~\ref{sec:SimpleDiscretization}.
This would render the kinetic-energy matrix significantly more sparse at the
expense of sacrificing the exact continuum dispersion relation.
In order to reach very large volumes, this may be a good tradeoff.

\subsection{Results for few-body resonances}

In the two-particle sector it is well
established~\cite{Wiese:1988qy,Luscher:1991cf,Rummukainen:1995vs} that resonance
states are manifest as avoided level crossings in the volume dependent energy
spectrum $E_i(L)$, where $i$ is an index labeling the discrete states in the
box.
Ref.~\cite{Klos:2018sen} found that shifted Gaussian potentials,
\begin{align}
 V(r)=V_0 \exp \biggl(-\Bigl(\frac{r-a}{R_0}\Bigr)^2\biggr) \,,
\label{eq:shiftedGauss}
\end{align}
are well suited to generate narrow resonances without much need to fine tune the
parameters of the potential and furthermore showed that fitting the inflection
points of individual level curves which form a plateau near the avoided crossing
gives excellent agreement for the resonance energies $E_R$ with determinations
from scattering phase shifts or from direct determinations as S-matrix poles on
the unphysical energy sheet.

This method also works well for three-particle systems, as established in
Ref.~\cite{Klos:2018sen} by comparing to a known case from the literature.
For three identical spin-0 bosons with mass $m=939.0~\MeV$ (mimicking neutrons)
interacting \via the two-body potential,
\begin{equation}
  V(r) = V_0 \exp\biggl(-\Bigl(\frac{r}{R_0}\Bigr)^2\biggr) + V_1
 \exp\biggl(-\Bigl(\frac{r-a}{R_1}\Bigr)^2\biggr) \,,
\label{eq:potblandon}
\end{equation}
where $V_0=-55~\MeV$, $V_1=1.5~\MeV$, $R_0=\sqrt{5}~\fm$, $R_1=10~\fm$, and
$a=5~\fm$, it is known that a resonance state exists at
$E_R=-5.31~\MeV$, with a half width of $0.12~\MeV$~\cite{Blandon:2007aa},
in addition to a two-body bound state at
$E=-6.76~\MeV$~\cite{Fedorov:2003jx} and a
three-boson bound state at $E=-37.35~\MeV$~\cite{Blandon:2007aa}.
Ref.~\cite{Fedorov:2003jx} obtained $E=-37.22~\MeV$ for this bound state
and $E_R=-5.96~\MeV$ and $\Gamma/2=0.40~\MeV$ for the three-body resonance,
which can be understood by noting that in this calculation the
potential~\eqref{eq:potblandon} was truncated to relative $S$ waves between
pairs~\cite{BB-20190905}.

\begin{figure}[tbp]
 \centering
 \includegraphics[width=0.75\columnwidth]{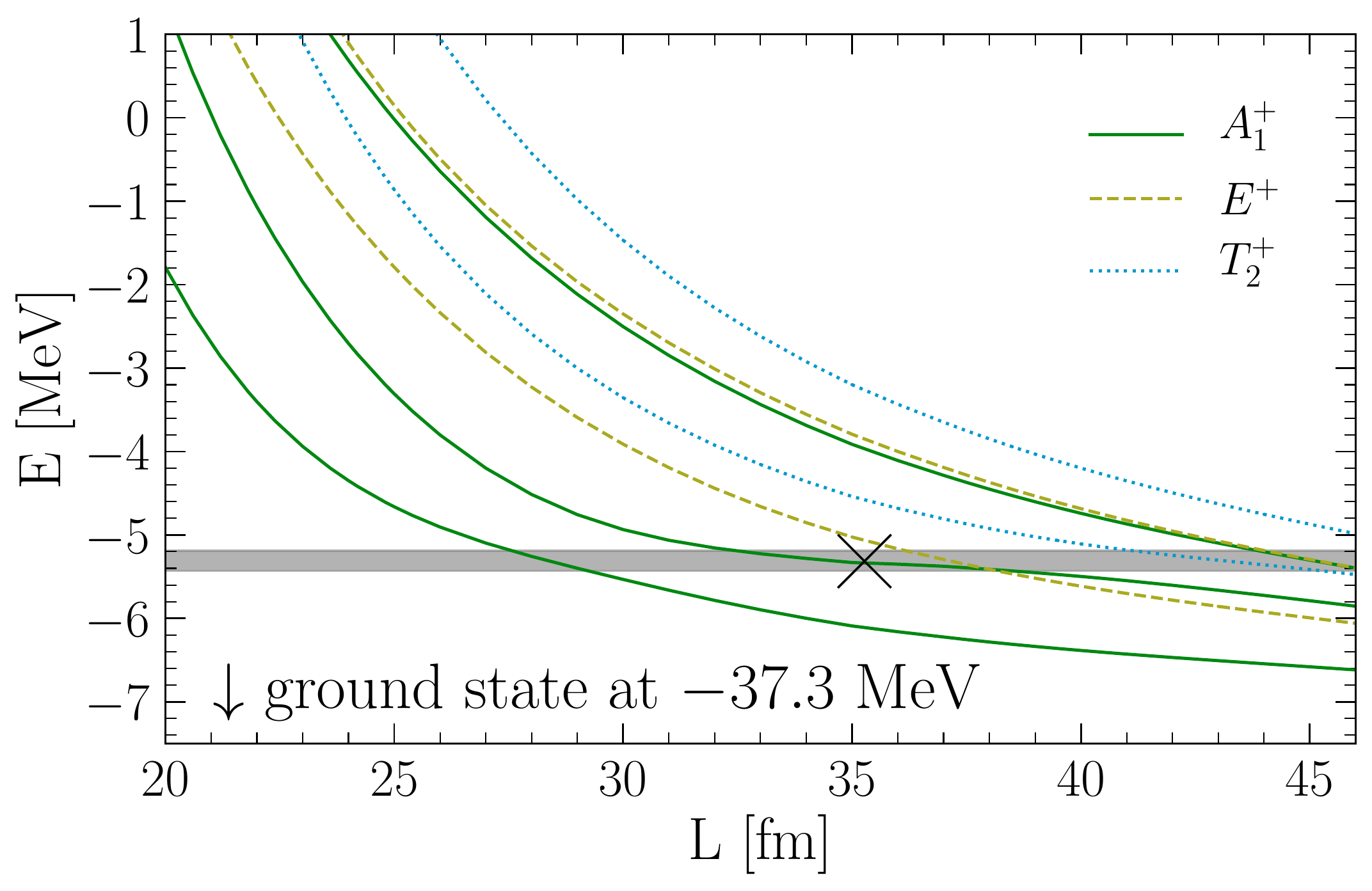}
 \caption{%
Energy spectrum of three bosons in finite volume for different box sizes $L$
interacting via the potential given in Eq.~\eqref{eq:potblandon}.
States corresponding to the irreducible representation $A_1$ of the cubic
symmetry group are shown as solid lines, whereas $E^+$ and $T_2^+$ states are
indicated as dashed and dotted lines, respectively.
The shaded area indicates the resonance position and width as calculated in
Ref.~\cite{Blandon:2007aa}, whereas the cross marks the inflection point used
here to extract the resonance energy (see text).
}
\label{fig:En-3b-Blandon-1-Pp}
\end{figure}

Using Eq.~\eqref{eq:potblandon} in a DVR calculation gives $E=-6.756(1)$ and
$E=-37.30(5)$ for the two- and three-boson ground states, respectively, in good
agreement with the results of Refs.~\cite{Fedorov:2003jx,Blandon:2007aa}.
These states both exhibit the exponential volume dependence as discussed in
Sec.~\ref{sec:BoundStates}.
In order to look for the three-boson resonance, the positive-parity three-body
spectrum is calculated as a function of $L$, shown in
Fig.~\ref{fig:En-3b-Blandon-1-Pp}.
These calculations used $n=26$ DVR points at smaller volumes, and up to $n=30$
for box sizes $L\sim 40~\fm$ to obtain sufficiently converged results.
Figure~\ref{fig:En-3b-Blandon-1-Pp} furthermore indicates the cubic-group
irreducible representations that individual levels belong to.
These assignments were determined by running a set of cubic-projected
calculations at selected volumes, while for computational efficiency the bulk of
the calculation did not use the cubic projection.
In general, it suffices to check the symmetry before and after each crossing
observed in the spectrum to determine whether or not it is an avoided crossing
(which only occurs between levels with the same quantum numbers) or an actual
one.

The levels corresponding to $A_1^+$ states show an avoided
crossing at about the expected resonance energy from Ref.~\cite{Blandon:2007aa},
which is indicated in Fig.~\ref{fig:En-3b-Blandon-1-Pp} as a shaded horizontal
band, the width of which corresponds to $E_R\pm\Gamma/2$.
For the other states (with quantum numbers $E^+$ and $T_2^+$) shown in the
figure we do not observe avoided crossings or plateaus.
At $L\sim 38~\fm$ there is an actual crossing between $A_1^+$ and an $E^+$
levels.
The resonance energy is extracted from the avoided crossing by the
inflection-point method, fitting polynomials
\begin{equation}
 E(L) = \sum_{k=0}^{k_\text{max}} c_k L^k \,,
\label{eq:fitfunction}
\end{equation}
to the $A_1^+$ curves participating in the avoided crossing.
This procedure gives $E_R = -5.32(1)~\MeV$, with the uncertainty stemming from
the fact that the lower level, which does not exhibit a very pronounced plateau
shape, does not constrain the fit very well.
The combined result from both energy levels however gives excellent agreement
with the resonance energy determined in Ref.~\cite{Blandon:2007aa}.
This is a clear indication that the method gives robust access to few-body
resonance energies.

\begin{figure*}[bhtp]
\centering
\includegraphics[width=0.66\textwidth]{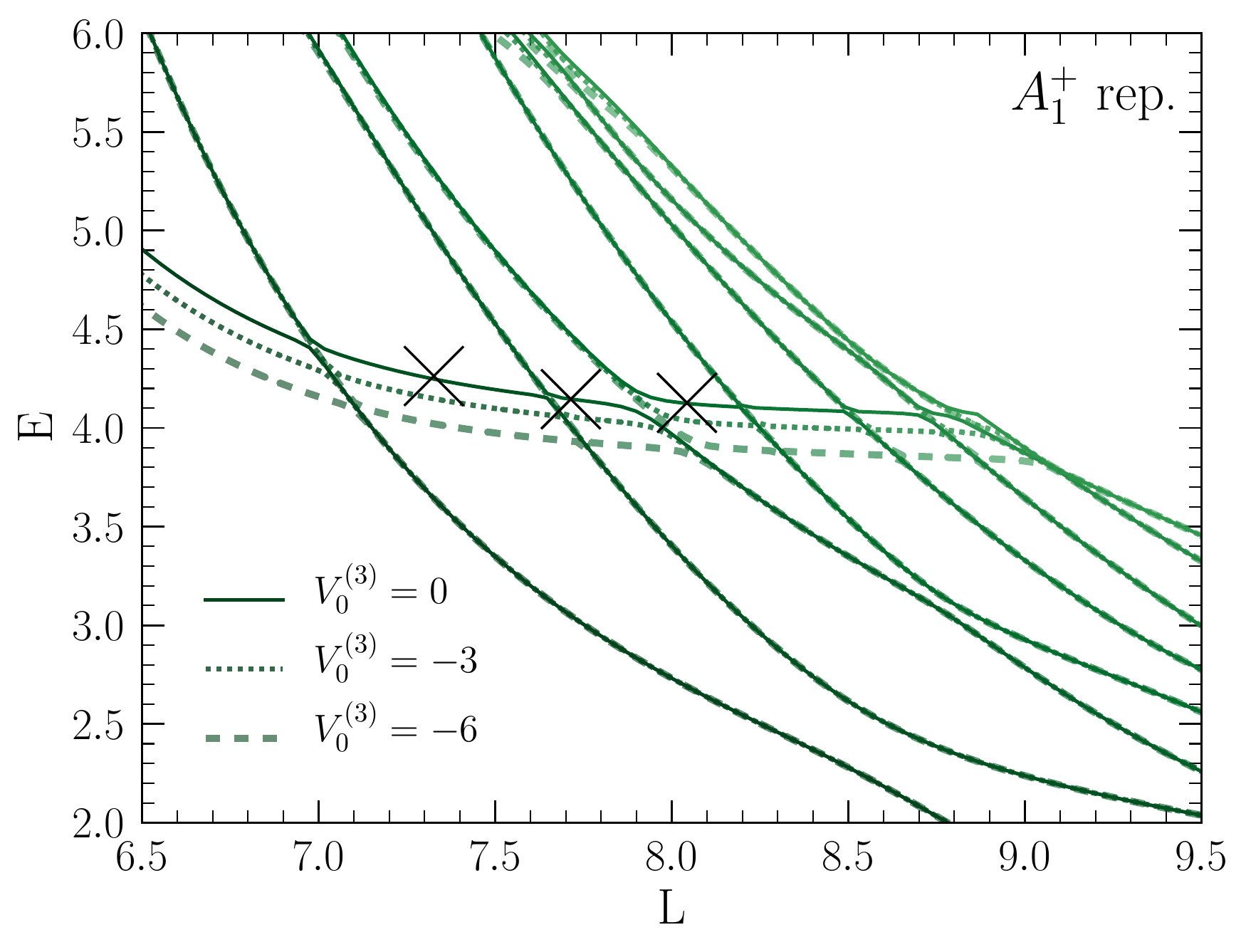}
\caption{Energy spectrum of three bosons in finite volume for different box
sizes $L$.
The solid line shows the spectrum for three bosons  interacting purely via the
shifted Gaussian potential given in Eq.~\eqref{eq:shiftedGauss} with $V_0=2.0$
while the dashed and dotted lines show results with an additional attractive
three-body force as in Eq.~\eqref{eq:V3}.
With increasing three-body force the avoided level crossing is shifted to lower
energy, while the rest of the spectrum remains unaffected.
}
\label{fig:En-3b-Dietz2p0-A1p}
\end{figure*}

Finding good agreement with Ref.~\cite{Blandon:2007aa} establishes the validity
and quantitative accuracy of the finite-volume method to extract three-body
resonances.
However, given that that the potential~\eqref{eq:potblandon} supports a two-body
bound state, and moreover the closeness of the resonance energy to that
threshold, the three-body resonance it generates should be considered an
effective two-body phenomenon.
To assess the method for the discovery of ``genuine'' three-body resonance, \ie,
states with no two-body decay channel, it is useful to consider a shifted
Gaussian potential as given in Eq.~\eqref{eq:shiftedGauss}, which does not
support any two-body bound states.
A three-body spectrum for this case, using $V_0=2.0$, $a=3$, and $R_0=1.5$,
is shown in Fig.~\ref{fig:En-3b-Dietz2p0-A1p}.
This spectrum, completely projected onto $A_1^+$ quantum numbers, features a
pronounced sequence of avoided level crossings between $E=4.0$ and $E=4.5$.
Using the same inflection-point fit
method as discussed above, one extracts $E_R = 4.18(8)$ as a potential resonance
energy by using the three points marked with crosses in
Fig.~\ref{fig:En-3b-Dietz2p0-A1p}.
In addition to this, there are several avoided crossings at lower
energies that have a significant slope with respect to changes in the box size.
These are interpreted as two-body resonances---known to exist at $E_{R} \sim
1.6$ for this potential~\cite{Klos:2018sen}---embedded into the three-body
spectrum.
This hypothesis can be validated by repeating the calculation with an added
short-range three-body force, as given in Eq.~\eqref{eq:V3}, setting
$R_0^{(3)}=1.0$ and varying $V_0^{(3)}$.
Using a set of negative values for $V_0^{(3)}$ (indicated in
Fig.~\ref{fig:En-3b-Dietz2p0-A1p}) leaves the lower avoided crossings (and in
fact most of the $L$-dependent spectrum) unaffected, whereas the upper plateau
set is moved downwards as $V_0^{(3)}$ is made more negative.
Since the range $R_0^{(3)}=1.0$ was chosen small (compared to the
box sizes considered), one should indeed expect it to primarily affect states
that are localized in the sense that their wavefunctions are confined to a
relatively small region in the finite volume.
Interpreting a resonance as a nearly bound state, its wavefunction should
satisfy this criterion, whereas three-body scattering states or states where
only two particles are bound or resonant are expected to have a large spatial
extent.
This intuitive picture gives confidence that indeed a genuine three-body
resonance is seen in Fig.~\ref{fig:En-3b-Dietz2p0-A1p}.

\begin{figure*}[tb]
\centering
\begin{minipage}{0.5\textwidth}
\centering
\includegraphics[width=0.99\columnwidth]{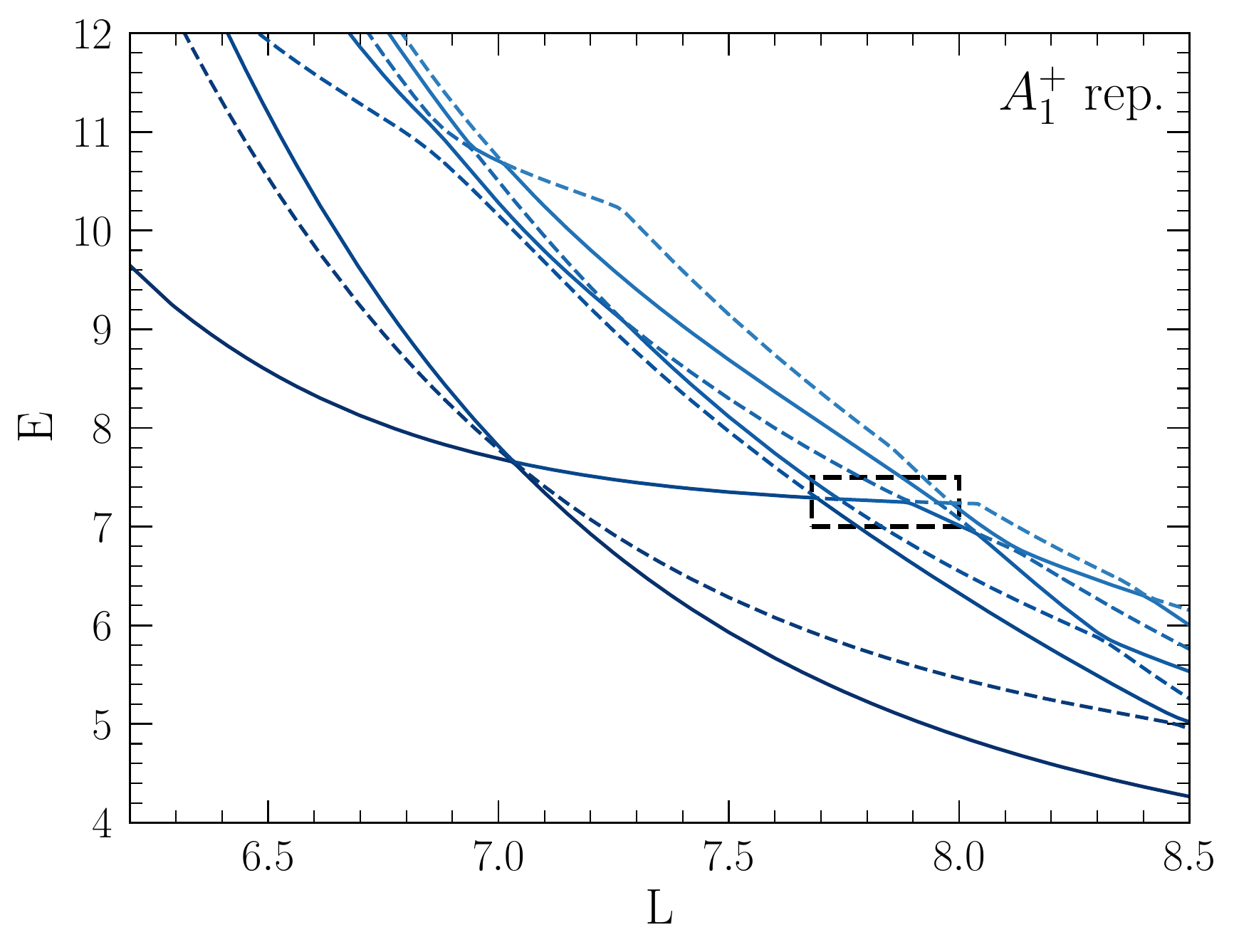}
\end{minipage}\begin{minipage}{0.5\textwidth}
\centering
\includegraphics[width=0.99\columnwidth]{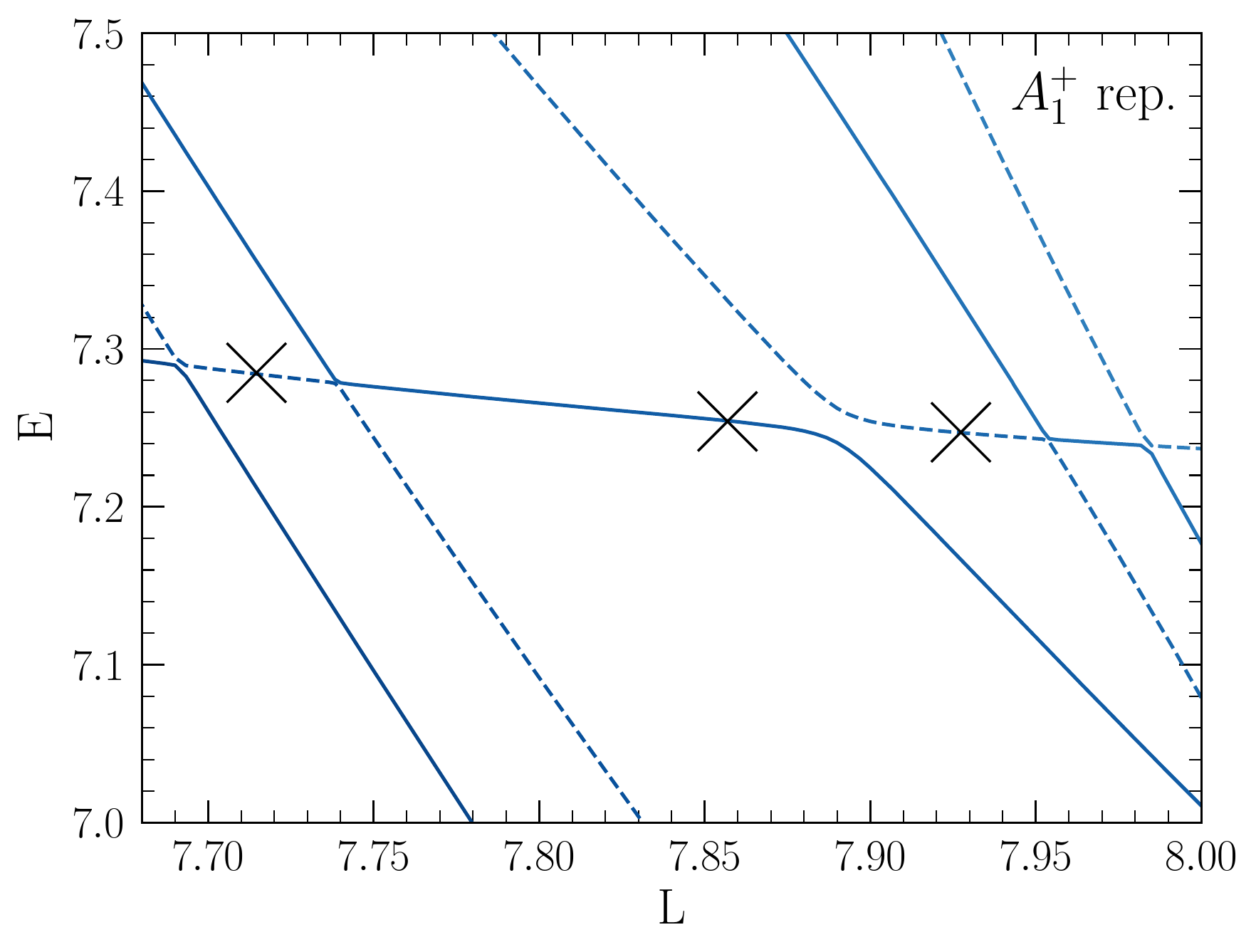}
\end{minipage}
\caption{Energy spectrum of four bosons in finite volume for
different box sizes $L$ interacting \via the shifted Gaussian potential given in
Eq.~\eqref{eq:shiftedGauss} with $V_0=2.0$.
The dashed rectangle in the left panel indicates the zoomed region shown in the
right panel.
All crossings are avoided because the spectrum is fully projected on states with
the same quantum numbers.
The crosses mark the inflection points used to extract the resonance energy (see
text).}
\label{fig:En-4b-Dietz2p0-A1p}
\end{figure*}
\begin{figure}[bt]
\centering
\includegraphics[width=0.66\columnwidth]{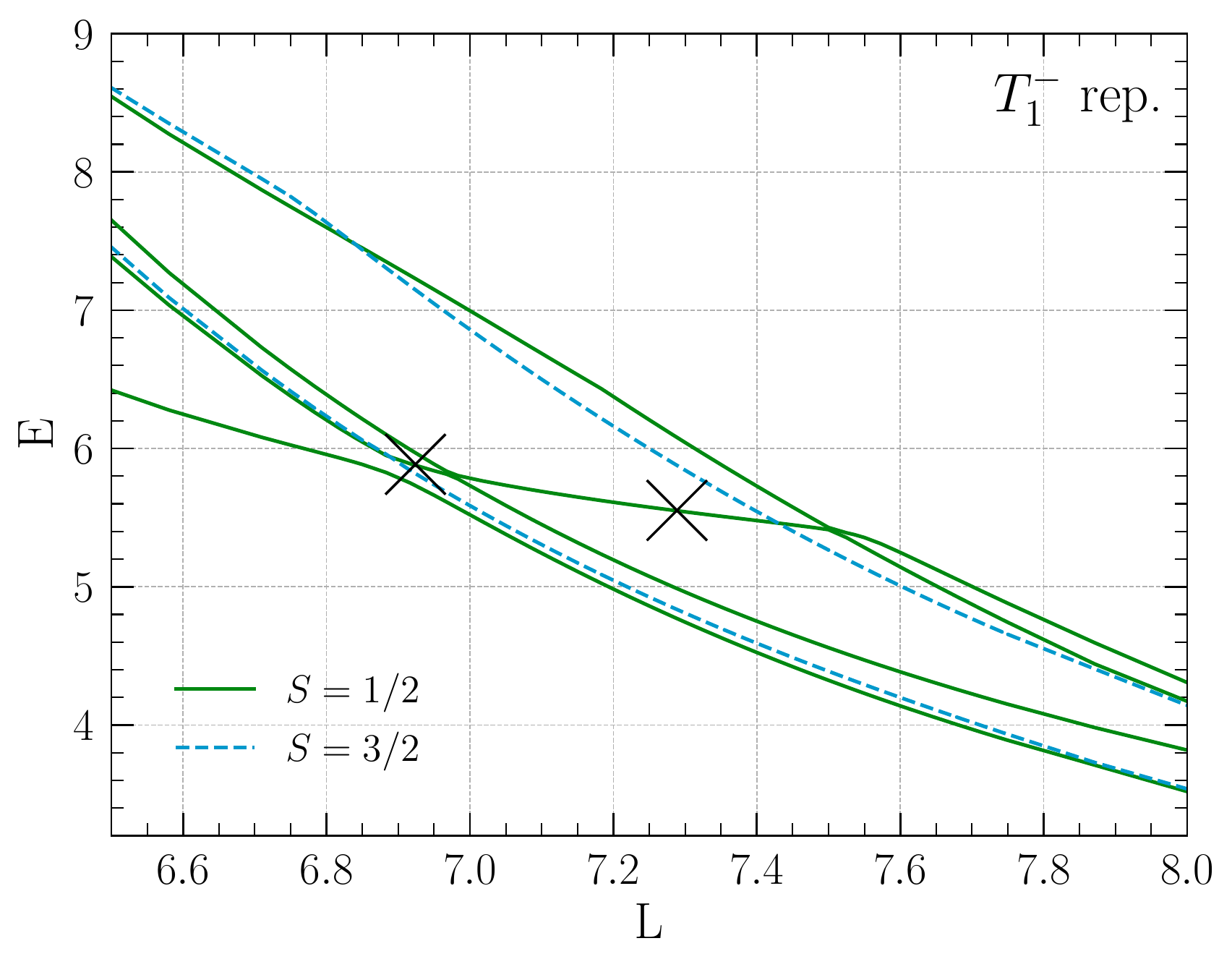}
\caption{Negative-parity energy spectrum of three fermions in finite volume for
different box sizes $L$ interacting \via the shifted Gaussian potential given in
Eq.~\eqref{eq:shiftedGauss} with $V_0=2.0$.
All levels shown in the plot were found to belong to the $T_1^-$ cubic
representation by performing fully projected calculations at selected volumes.
Results are shown in the spin $S=1/2$ and $S=3/2$ channels.
The crosses mark the inflection points used to extract the resonance energy (see
text).}
\label{fig:En-3f-Dietz2p0-Pm}
\end{figure}

Similar features are found for calculations with the same shifted Gaussian
potential of four-boson and three-fermion systems.
These are shown, respectively, in Figs.~\ref{fig:En-4b-Dietz2p0-A1p}
and~\ref{fig:En-3f-Dietz2p0-Pm}, with resonance energies extracted as
$E_R^{\text{4b}}=7.26(2)$ and $E_R^{\text{3f}}=5.7(2)$.
For the three-fermion calculation one needs to take into account that the
overall antisymmetry of the wavefunction can be realized \via different
combinations of spin and spatial parts.
For negative parity one finds the six lowest levels, shown in
Fig.~\ref{fig:En-3f-Dietz2p0-Pm}, to all belong to the $T_1^-$ cubic
representation, which in this case has been determined by running calculations
with full cubic projections at selected volumes while otherwise only restricting
the overall parity.
Since the interaction is spin independent, total angular momentum $\ell$ and
spin $S$ are separately good quantum numbers in infinite volume, and in the
finite volume one likewise has $\Gamma$ and $S$ to characterize states.
The latter, which can be $S=1/2$ or $S=3/2$ for three spin-$1/2$ fermions, is
determined by
running calculations with fixed spin $z$-component at selected volumes, which
can be realized by restricting the set of DVR basis states.
Since $S=3/2$ states show up with both $S_z=3/2$ and $S_z=1/2$, whereas
$S=1/2$ states are absent for $S_z=3/2$, one finds that four of the six levels
shown in Fig.~\ref{fig:En-3f-Dietz2p0-Pm} have $S=1/2$, whereas the other two
(given by the dashed lines in Fig.~\ref{fig:En-3f-Dietz2p0-Pm}) have $S=3/2$.
The resonance signature is found for $S=1/2$ in this case.

\section{Summary and outlook}
\label{sec:Summary}

Finite-volume calculations provide an intriguing way to study physical systems.
Their infinite-volume properties are encoded in the response of discrete energy
levels to variations in the size of the volume: it is the physical S-matrix what
governs the precise form of the volume dependence and therefore by studying the
latter one can infer properties of the former.

For bound states this relation is manifest as asymptotic wavefunctions,
characterized by their exponential fall-off scale and asymptotic normalization
constants (which by analyticity are related to the S-matrix), determining the
volume dependence in a direct manner.
Knowing the precise form of this dependence
enables controlled extrapolations of $N$-body states from small to infinite
volume.
This is relevant in nuclear physics for Lattice
QCD~\cite{Nicholson:2015pys,Berkowitz:2015eaa,Yamazaki:2015asa,%
Yamazaki:2015vjn,Inoue:2014ipa,Etminan:2014tya,Beane:2014ora,Chang:2015qxa,%
Savage:2016kon} as well as Lattice
EFT~\cite{Epelbaum:2013paa,Elhatisari:2015iga,Elhatisari:2016owd} calculations,
and more broadly for finite-volume calculations of for example bound cold
atomic systems.
Furthermore, the finite-volume relations provide a direct way to calculate
asymptotic normalization coefficients, which play an important role for
low-energy capture processes in nuclear astrophysics and are notoriously
difficult to determine experimentally.
Since most of these reactions involve charged particles, future work which
extends the relations to include the long-range Coulomb
interaction will open the door to many interesting applications.
Beyond this, studying bound states for which the nearest breakup threshold
involves a splitting into more than two clusters requires more research in order
to understand the additional power-law factors that arise from continuum effects
and are so far known only for a few specific
cases~\cite{Konig:2017krd,Meissner:2014dea}.

Resonances are found to be robustly manifest as avoided level crossing in the
spectrum as the size of the box is varied.
The work discussed here establishes that this methods is able, at a quantitative
level, to extract few-body resonance energies and therefore provides a discovery
tool for states which are otherwise very difficult to tackle.
This is important in light of much disagreement in the literature regarding the
possibility of three- and four-neutron resonances from a theoretical
perspective~\cite{Witala:1999pm,Lazauskas:2005ig,Hiyama:2016nwn,Klos:2016fdb,%
Shirokov:2016ywq,Gandolfi:2016bth,Fossez:2016dch,Deltuva:2018lug,%
Deltuva:2019mnv}.
While such determinations are made difficult by the fact that conjectured
resonance states have supposedly very small energies---requiring converged DVR
calculations in large boxes due to the power-law behavior the finite-volume
energy levels---carrying out such calculations can provide valuable insights
regarding the existence of such exotic nuclear states.
Apart from that, the finite-volume technique provides an interesting and
conceptually straightforward way to study other resonances as well, such as for
example the Hoyle state in \isotope[12]{C}, or metastable states in cold atomic
systems.
While the inflection-point method discussed here seems to robustly capture the
real part $E_R$ of the overall resonance position, more formal developments are
necessary to extract resonance width from the details of the spectrum.
There is some interesting recent work in this
direction~\cite{Romero-Lopez:2019qrt}.
Overall, there are many exciting opportunities for future research.

\begin{acknowledgement}
I would like to thank Hans-Werner Hammer, Philipp Klos, Dean Lee, Joel Lynn and
Achim Schwenk for the collaboration that led to the original works summarized in
this paper.
I am furthermore grateful to Dean Lee for pointing me to the connection between
the plane-wave DVR and the simple finite-differences discretization.
This work was supported in part by the ERC Grant No.\ 307986 STRONGINT and the
Deutsche Forschungsgesellschaft (DFG) under Grant SFB 1245.
The numerical computations were performed on the Lichtenberg high performance
computer of the TU Darmstadt and at the Jülich Supercomputing Center.
\end{acknowledgement}

\bibliographystyle{spphys}

\end{document}